\documentclass{aa}
\usepackage{txfonts}
\usepackage{graphicx}
\usepackage{natbib}
\usepackage{amssymb}
\usepackage{color}
\bibpunct{(}{)}{;}{a}{}{,}
\begin{document}
\title{Circumstellar molecular line emission from S-type AGB stars: \\
 Mass-loss rates and SiO abundances}
\author{S. Ramstedt\inst{1} \and F. L. Sch\"oier\inst{1,2} \and H. Olofsson\inst{1,2}}
\offprints{S. Ramstedt}
\institute{Department of Astronomy, Stockholm University, AlbaNova University Center, SE-106 91 Stockholm, Sweden \\ \email{sofia@astro.su.se} \and Onsala Space Observatory, SE-439 92 Onsala, Sweden}    
\date{Received; accepted}
\abstract{}{The main aim is to derive reliable mass-loss rates and circumstellar SiO abundances for a sample of 40 S-type AGB stars based on new multi-transitional CO and SiO radio line observations. In addition, the results are compared to previous results for M-type AGB stars and carbon stars to look for trends with chemical type.}{The circumstellar envelopes are assumed to be spherically symmetric and formed by a constant mass-loss rate. The mass-loss rates are estimated from fitting the CO observations using a non-local, non-LTE radiative transfer code based on the Monte-Carlo method. In the excitation analysis, the energy balance equation is solved self-consistently simultaneously as the radiative transfer and the temperature structure of the gas is derived. Effects of dust grains are also included in the molecular excitation analysis. Once the physical properties of the circumstellar envelopes are determined, the same radiative transfer code is used to model the observed SiO lines in order to derive circumstellar abundances and the sizes of the SiO line-emitting regions.}{We have estimated mass-loss rates of 40 S-type AGB stars and find that the derived mass-loss rates have a distribution that resembles those previously derived for similar samples of M-type AGB stars and carbon stars. The estimated mass-loss rates also correlate well with the corresponding expansion velocity of the envelope, in accordance with results for M-type AGB stars and carbon stars. In all, this indicates that the mass loss is driven by the same mechanism in all three chemical types of AGB stars. In addition, we have estimated the circumstellar fractional abundance of SiO relative to H$_{2}$ in 26 of the sample S-type AGB stars. The derived SiO abundances are, on average, about an order of magnitude higher than predicted by stellar atmosphere thermal equilibrium chemistry, indicating that non-equilibrium chemical processes determines the abundance of SiO in the circumstellar envelope. Moreover, a comparison with the results for M-type AGB stars and carbon stars show that for a certain mass-loss rate, the circumstellar SiO abundance seems independent (although with a large scatter) of the C/O-ratio. }{In our comparison of S-type AGB stars with carbon stars and M-type AGB stars, we find no large differences in circumstellar physical properties or SiO abundances depending on the chemical type of the star.}   
   
\maketitle
  

\section{Introduction}
\label{s:intro}
The final evolutionary stage of low- to intermediate-mass stars, as they ascend the asymptotic giant branch (AGB), is characterized by an intense mass loss. The stellar wind builds up gradually and creates a circumstellar envelope (CSE), carrying gas and dust from the star into the interstellar medium. 
The molecular setup and grain types in CSEs are to a large extent determined by the C/O-ratio in the photosphere of the central star. The AGB stars are normally divided into two distinct spectral types: M-type stars, with C/O$<$1 and carbon stars, with C/O$>$1. Abundance analysis has also revealed stars with photospheric C/O-ratios close to unity (within approximately 5 \%), so called S-type AGB stars \citep{scalross76}. The spectra of S-type AGB stars are dominated by ZrO bands (as opposed to spectra of M-type AGB stars which are dominated by TiO bands) indicating that the S-type AGB stars have an enhancement in elements formed through the slow neutron capture process. 

Due to having a C/O-ratio close to unity, it is tempting to identify S-type AGB stars with a brief transitional phase as the star evolves from an oxygen-rich M-type AGB star into a carbon star. Dredge-up of carbon from He-shell burning would change the spectral type of the star sequentially: M-MS-S-SC-C. As possible transition objects, S-type AGB stars might very well help to achieve a deeper understanding of the chemical evolution as a star ascends the AGB, as well as shed light on the mass-loss mechanism(s), which is (are) not yet fully understood in detail. Several surveys of CO emission from their CSEs have been performed \citep{bieglatt94,sahaliec95,biegetal98,groedejo98,joriknap98,ramsetal06}. Circumstellar molecular line emission from molecules other than CO has previously been searched for, and detected, only in a handful of objects and only for HCN and SiO \citep{bieglatt94,biegetal98}. 

The aim of this work is to thoroughly investigate the circumstellar physical and chemical properties of a sample of S-type AGB stars. The physical properties of the CSEs, in particular the mass-loss rates that produced them, are determined from the CO data using detailed, non-LTE, radiative transfer modelling which self-consistently calculates also the gas kinetic temperature. Detailed studies of the mass-loss properties of samples of carbon stars \citep{schoolof01,schoetal02}, M-type AGB stars \citep{olofetal02,delgetal03}, and S-type AGB stars \citep{ramsetal06} have previously been performed. 

Lately, the mechanisms driving mass loss in M-type AGB stars have been much debated \citep{woit06,hofnande07,hoef08}, and for the S-type AGB stars this is a matter of a long-standing debate \citep[e.g.,][]{willdejo88,sahaliec95}. The supposed lack of any free oxygen and/or free carbon to form dust (due to a C/O-ratio close to unity) has lead to the suggestion that the S-type AGB stars would not be able to drive a wind as efficiently as the M-type AGB stars and the carbon stars. However, the results of \citet{ramsetal06} show no apparent differences between the mass-loss rate distribution of the S-type AGB stars and those found for carbon and M-type AGB stars.

Once the physical properties are known, from the CO analysis, they can be used to estimate abundances of other molecules in the CSE. The analysis of the S-type AGB stars in \citet{ramsetal06} was based on observations of CO($J$\,=\,3$\rightarrow$\,2) data gathered at the APEX telescope supplemented with data collected from the literature. When examining the published data, \citet{ramsetal06} found a rather large scatter in the reported line intensities [especially for CO($J$\,=\,2$\rightarrow$\,1) data] for individual objects, even when observed with same telescope. Since then, we have obtained more observational data, which warrants a re-analysis of the \citet{ramsetal06} work, and the results are reported here. 

We have, as a first step after the CO observations, searched for circumstellar SiO radio line emission in several rotational transitions. Using the physical structure of the CSEs derived from the CO modelling, we estimate the abundance of SiO in the same sample of S-type AGB stars based on a detailed excitation analysis. There exists strong evidence that the circumstellar SiO emission carries information on the region where the mass loss is initiated and where the dust formation takes place \citep{reidmora81,reidment97,schoetal06a} making it a particularly interesting molecule to study. \citet{schoetal06} modelled circumstellar SiO line observations from a sample of carbon stars and when comparing this to a similar analysis performed by \citet{delgetal03} for a large sample of M-type AGB stars, they found no apparent distinction between their circumstellar SiO abundance distributions. For the carbon stars, the derived abundances are several orders of magnitude higher than expected from thermal equilibrium stellar atmosphere chemistry. A possible explanation for the high SiO abundances derived for the carbon stars is the influence of a shock chemistry in the inner part of the wind \citep{cher06}. With the analysis performed in this work, the S-type AGB stars can be added to this comparison (see Sect.~\ref{ss:dis:sio}). 

The sample of S-type AGB stars and the observational data are presented in Sects~\ref{s:sample} and \ref{s:obsdata}. The radiative transfer modelling is described in Sect.~\ref{s:COmod}. The results are given and discussed in Sects~\ref{s:res} and \ref{s:discuss}, respectively. Finally, the conclusions are given in Sect.~\ref{s:conc}.


\section{The sample}
\label{s:sample}
\begin{table}
\caption{The sample with variable types, periods $P$, luminosities $L_{\star}$, and distances $D$.}
\label{sample}
$$
\begin{array}{p{0.3\linewidth}lccccc}
\hline
\noalign{\smallskip}
\multicolumn{1}{l}{{\mathrm{Source}}} &
\multicolumn{1}{l}{{\mathrm{Var.}}} &&
\multicolumn{1}{c}{{\mathrm{P}}} &
\multicolumn{1}{c}{L_{\star}} && 
 \multicolumn{1}{c}{D}\\ 
&
\multicolumn{1}{l}{\mathrm{type}} &&
\multicolumn{1}{c}{[\mathrm{days}]} &
\multicolumn{1}{c}{[\mathrm{L}_\odot]} &&
\multicolumn{1}{c}{[\mathrm{pc}]} 
\\
\noalign{\smallskip}
\hline
\noalign{\smallskip}
\object{R And} & \rm{M} &&  409 & 6000 && \phantom{1}300 \\
\object{W And} & \rm{M} && 397 & 5800 && \phantom{1}280 \\
\object{Z Ant} & \rm{SR} && 104 & 4000 && \phantom{1}470 \\
\object{VX Aql} & \rm{M:} && \cdots & 4000 && \phantom{1}790 \\
\object{W Aql} & \rm{M} && 490 & 6800 && \phantom{1}230 \\	
\object{AA Cam} & \rm{Lb} && \cdots & 4000 && \phantom{1}800 \\
\object{T Cam} & \rm{M} && 374 & 5600 && \phantom{1}540 \\
\object{S Cas} & \rm{M} && 611 & 8000 && \phantom{1}440 \\	
\object{V365 Cas}	& \rm{SRb} && 136 & 4000 && \phantom{1}625 \\
\object{WY Cas}	& \rm{M} && 477 & 6700 && \phantom{1}600 \\
\object{AM Cen} & \rm{Lb} && \cdots & 4000 && \phantom{1}750 \\
\object{TT Cen} & \rm{M} && 462 & 6500 && \phantom{1}880 \\
\object{UY Cen} & \rm{SR} && 115 & 4000 && \phantom{1}590 \\
\object{V386 Cep} & \rm{SR} && \cdots & 4000 && \phantom{1}470 \\
\object{T Cet} & \rm{SRb} && 159 & 4000 && \phantom{1}240 \\	
\object{AA Cyg} & \rm{SRb} && 213 & 4000 && \phantom{1}480 \\
\object{AD Cyg} & \rm{Lb} && \cdots & 4000 && \phantom{1}980 \\
\object{R Cyg} & \rm{M} && 426 & 6100 && \phantom{1}440 \\	
\object{$\chi$ Cyg}	& \rm{M} && 407 & 5900 && \phantom{1}110 \\
\object{TV Dra}	& \rm{Lb} && \cdots & 4000 && \phantom{1}390 \\
\object{DY Gem} & \rm{SRa} && \cdots & 4000 && \phantom{1}680 \\
\object{R Gem} & \mathrm{M} && 370 & 5500 && \phantom{1}710 \\
\object{ST Her} & \rm{SRb} && 148 & 4000 && \phantom{1}300 \\
\object{RX Lac} & \rm{SRb} && 174 & 4000 && \phantom{1}310 \\
\object{GI Lup} & \rm{M} && 325 & 5000 && \phantom{1}690 \\	
\object{R Lyn} & \rm{M} && 379 & 5600 && \phantom{1}850 \\
\object{Y Lyn} & \rm{SRc} && 110 & 4000 && \phantom{1}260 \\
\object{S Lyr} & \rm{M} && 438 & 6300 && 1210 \\	
\object{FU Mon}	& \rm{SR} && 310 & 4000 && \phantom{1}780 \\
\object{RZ Peg}	& \rm{M} && 439 & 6300 && \phantom{1}970 \\
\object{RT Sco} & \rm{M} && 449 & 6400 && \phantom{1}270 \\
\object{ST Sco} & \rm{SRa} && 195 & 4000 && \phantom{1}380 \\
\object{RZ Sgr} & \rm{SRb} && 223 & 4000 && \phantom{1}730 \\
\object{ST Sgr} & \rm{M} && 395 & 5800 && \phantom{1}540 \\
\object{T Sgr} & \rm{M} && 392 & 5800 && \phantom{1}590 \\	
\object{EP Vul} & \rm{Lb} && \cdots & 4000 && \phantom{1}510 \\
\object{DK Vul} & \rm{SRa} && 370 & 4000 && \phantom{1}750 \\
\object{AFGL 2425} & \rm{U} && \cdots & 4000 && \phantom{1}610 \\
\object{CSS2 41} & \rm{U} && \cdots & 4000 && \phantom{1}880 \\	
\object{IRC--10401}	& \rm{U} && \cdots & 4000 && \phantom{1}430 \\
\noalign{\smallskip}
\hline
\end{array}
$$
\end{table}

\citet{joriknap98} presents a sample of 124 S-type stars from the list of \citet{chenetal95} providing cross identifications between the {\em General Catalogue of Galactic S stars}, the IRAS {\em Point Source Catalogue} (PSC), and the {\em Guide Star Catalogue}. Only stars having flux densities of good quality in the 12, 25, and 60 $\mu$m bands in the IRAS PSC were retained in their sample. We have chosen the stars in the sample of Jorissen \& Knapp that have previously been detected in circumstellar CO emission. Five stars have been added from the samples of \citet{groedejo98} and \citet{sahaliec95} [selected from the S-type catalogs of \citet{step84,step90}], and one star was added from the sample of \citet{bieglatt94} [selected from the list of \citet{jura88}]. These six stars have also been detected previously in circumstellar CO emission. Only S-type stars with Tc lines detected in their spectra and with detectable infrared excess are safely identified as intrinsic thermally pulsing AGB stars. S-type stars with no Tc lines and no infrared excess are most likely extrinsic S-stars, i.e., they are part of binary systems and their chemical peculiarities are due to mass transfer across the system. Jorissen \& Knapp investigate the extrinsic/intrinsic characteristics of their sample and we have only retained stars which they classify as intrinsic. The stars from the Groenewegen \& de Jong and Sahai \& Liechti samples have all been detected in Tc and/or show infrared excess. We therefore believe that our sample only consists of intrinsic S-type AGB stars. 
Our sample of 40 stars is listed in Table~\ref{sample}. Our selection criteria might introduce a bias in the sample toward S-type stars that have higher mass-loss rates, and S-type stars without or with very little mass loss will be missed. The reason for this selection is to be able to detect line emission from other molecules than CO. 

\begin{figure}[h]
\raggedright 
{\includegraphics[width=\columnwidth]{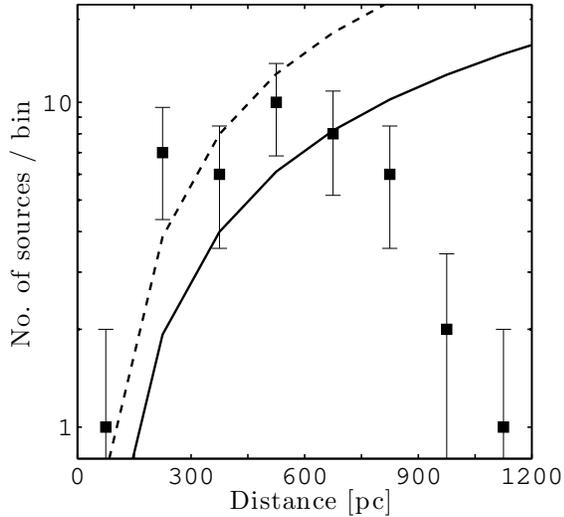}}
\caption{The distance distribution of our sample of S-type AGB stars (squares) overlayed with the distribution found in \citet{jura90} assuming a smooth surface distribution (solid line, see text in Sect.~\ref{s:sample}) of S-type AGB stars in the solar neighbourhood. The dashed line shows a fit to the observational results. The errors are based on Poisson statistics.}
\label{complete}
\end{figure}

For the Miras (M), bolometric luminosities are estimated using the period-luminosity relation of \citet{whitetal94}, and periods are taken from the {\em General Catalogue of Variable Stars} \citep{kholetal99}. For the semiregular (SR, SRa, and SRb) and irregular variables (Lb), a luminosity of $4000\, L_{\odot}$ was assumed in accordance with \citet{olofetal02}. Corrections for the interstellar extinction, based on their galactic latitude, were applied for each star \citep{groeetal92}. Distances are derived by fitting the spectral energy distribution (SED) calculated with DUSTY (Sect.~\ref{s:dustmod}), to observed fluxes (Sect.~\ref{s:flux}). The derived distances, and adopted periods and luminosities are presented in Table~\ref{sample}, together with the variability type. The derived distances from the SED modelling agree well with distances from Hipparcos parallaxes when available (for 7 stars).  

Fig.~\ref{complete} shows the distance distribution of our sample in 150 pc bins. The distance distribution is overlayed with a distribution assuming that the stars are evenly distributed, that the surface density of carbon stars is 40\,kpc$^{-2}$ with a scale height of 200\,pc, and that there are 1/3 as many S-type stars as carbon stars in the solar neighbourhood \citep{wingyork77,jura90}. A direct fit to our sample within 600 pc, gives a surface density of  20\,kpc$^{-2}$ with a scale height of approximately 150\,pc (dashed line, Fig.~\ref{complete}). Considering the uncertainties involved in the distance estimate, classification, etc., we find the results to be consistent and believe that our sample is representative of mass-losing S-type AGB stars and complete to about 600 pc. The sample of carbon stars studied by \citet{schoolof01} is representative of C-type AGB stars and complete to about 500 pc, while the completeness of the M-type AGB star samples of \citet{olofetal02} and \citet{delgetal03} have not been thoroughly investigated. 


\section{Observational data}
\label{s:obsdata}

When modelling circumstellar molecular line emission it is essential to observe as many different transitions as possible of the molecule under study. The derived mass-loss rate (when assuming a spherically symmetric, smooth wind) should be considered as the average mass-loss rate that created the CSE probed by the observed line emission. Different transitions probe slightly different regions in the CSE depending on their excitation requirement and thus the average is taken over a larger part of the CSE if more transitions are obtained. The same is true when estimating the radial abundance distribution and, in particular, when estimating the size of the emitting region of a specific molecule. In this work, we have tried to obtain observations of at least three transitions of CO and SiO for as many of the sample sources as possible (Tables~\ref{COintensities1}--\ref{SiOintensities2}).

\begin{table}
\caption{Energies of the upper levels, $E_{\rm{up}}$, main-beam efficiencies, $\eta_{\mathrm{mb}}$, and main-beam FWHM:s, $\theta_{\mathrm{mb}}$, at the frequencies of the observed CO and SiO lines.}
\label{eta_mb}
$$
\begin{array}{ccccccccccc}
\hline
\noalign{\smallskip}
\multicolumn{1}{c}{{\mathrm{Transition}}} &&
\multicolumn{1}{c}{{\mathrm{Frequency}}} &&
\multicolumn{1}{c}{{E_{\rm{up}}}} &&
\multicolumn{1}{c}{\eta_{\mathrm{mb}}}  && 
\multicolumn{1}{c}{\theta_{\mathrm{mb}}} &
\multicolumn{1}{c}{{\mathrm{Telescope}}} \\ 
&&
\multicolumn{1}{c}{{\mathrm{[GHz]}}} && 
\multicolumn{1}{c}{{\mathrm{[K]}}} && && 
\multicolumn{1}{c}{{\mathrm{['']}}} & \\
\noalign{\smallskip}
\hline
\noalign{\smallskip}
\mathrm{CO}(J=1\rightarrow0) && 115.271    && \phantom{1}5.5 && 0.43 && 33 & \mathrm{OSO} \\
\mathrm{CO}(J=1\rightarrow0) && 115.271    && \phantom{1}5.5 && 0.74 && 22 & \mathrm{IRAM} \\
\mathrm{CO}(J=2\rightarrow1) && 230.538    && 16.6 && 0.52 && 11 & \mathrm{IRAM} \\
\mathrm{CO}(J=3\rightarrow2) && 345.796    && 33.2 && 0.56 && 14 & \mathrm{JCMT} \\	
\mathrm{CO}(J=3\rightarrow2) && 345.796    && 33.2 && 0.70 && 18 & \mathrm{APEX} \\
&& && && & \\
\mathrm{SiO}(J=2\rightarrow1) && \phantom{1}86.847  && \phantom{1}6.3 && 0.65 && 44 & \mathrm{OSO} \\
\mathrm{SiO}(J=2\rightarrow1) && \phantom{1}86.847  && \phantom{1}6.3 && 0.78 && 29 & \mathrm{IRAM} \\
\mathrm{SiO}(J=5\rightarrow4) && 217.105    && 31.3 && 0.57 && 12 & \mathrm{IRAM} \\
\mathrm{SiO}(J=6\rightarrow5) && 260.518    && 43.8 && 0.69 && 21 & \mathrm{JCMT} \\
\mathrm{SiO}(J=8\rightarrow7) && 347.331    && 75.0 && 0.63 && 14 & \mathrm{JCMT} \\	
\mathrm{SiO}(J=8\rightarrow7) && 347.331    && 75.0 && 0.70 && 18 & \mathrm{APEX} \\
\noalign{\smallskip}
\hline
\end{array}
$$
\end{table}
%


\subsection{Observations of circumstellar CO and SiO radio line emission}
\label{ss:CO}

The first analysis of CO radio line emission from this sample was published in \citet{ramsetal06}. There, data from the literature (mainly $J$\,=\,1\,$\rightarrow$\,0 and 2\,$\rightarrow$1) was analysed together with $J$\,=\,1\,$\rightarrow$\,0 data collected with the Onsala Space Observatory (OSO) 20\,m telescope\footnote{The Onsala 20\,m telescope is operated by the  Swedish National Facility for Radio Astronomy, Onsala Space Observatory at Chalmers University of Technology, with support from the Swedish Research Council.} (January 2006), and $J$\,=\,3\,$\rightarrow$\,2 data of 18 sources collected with the APEX 12\,m telescope\footnote{The Atacama Pathfinder Experiment (APEX) is a collaboration between the Max-Plank-Institute f\"{u}r Radioastronomie, the European Southern Observatory, and the Onsala Space Observatory.} (from August to October 2005). 

The CO data base has since then been substantially extended. New observations of $J$\,=\,1\,$\rightarrow$\,0 and 2\,$\rightarrow$\,1 line emission were performed at the IRAM 30\,m telescope\footnote{IRAM is supported by CNRS/INSU (France), the MPG (Germany), and the IGN (Spain).} using the AB SIS receiver combination observing both polarizations simultaneously in August 2006. Further observations of $J$\,=\,3\,$\rightarrow$\,2 line emission were performed at APEX using APEX-2A during the autumn of 2006 and at the James Clerk Maxwell Telescope\footnote{The James Clerk Maxwell Telescope is operated by the Joint Astronomy Centre on behalf of the Science and Technology Facilities Council of the United Kingdom, the Netherlands Organisation for Scientific Research, and the National Research Council of Canada.} (JCMT) in April 2007. At the JCMT we used HARP due to problems with the B receiver. As spectrometers, the VESPA autocorrelator at IRAM, the FFTS at APEX, and the ACSIS digital autocorrelator at the JCMT were used.

New SiO observations of $J$\,=\,2\,$\rightarrow$\,1 line emission were performed at OSO using an SIS receiver and the autocorrelator spectrometer in the low-resolution mode in January 2006. Observations of $J$\,=\,2\,$\rightarrow$\,1 and 5\,$\rightarrow$\,4 line emission were performed at the IRAM 30\,m telescope in August 2006, and further observations of $J$\,=\,6\,$\rightarrow$\,5 and 8\,$\rightarrow$\,7 line emission were performed at JCMT during the autumn of 2006 and spring 2007. The observations at IRAM and the JCMT were obtained simultaneously as the CO observations using the same receiver and spectrometer setups. During 2006 and 2007 observations of $J$\,=\,8\,$\rightarrow$\,7 line emission was performed at APEX using the same setup as for the CO observations (all these lines are in the $v=0$ state). Simultaneously as the $J$\,=\,2\,$\rightarrow$\,1, $v=0$, observations at OSO, the $J$\,=\,2\,$\rightarrow$\,1, $v=1$ line (which is of maser origin) was observed. We observed SiO($J$\,=\,2$\rightarrow$1, $v$=1) line emission in 24 of our sample stars. For four of the stars, WY Cas, V386 Cep, TV Dra and R Gem, this is the first detection of SiO maser emission.

At OSO the observations were performed using the dual beam-switch mode. The observations at IRAM were performed using wobbler switching and a dual-polarization mode. At APEX position-switching was used with the reference position located at $+$2\arcmin~in azimuth. The JCMT observations were also performed using position-switching with a 2\arcmin~throw. Regular pointing checks were performed during all observations and typically found to be consistent within $\approx$3\arcsec of the pointing model.

The data was reduced using CLASS, Starlink, and XS\footnote{XS is a package developed by P. Bergman to reduce and analyze a large number of single-dish spectra. It is publicly available from {\tt ftp://yggdrasil.oso.chalmers.se}} by subtracting a first order polynomial baseline fitted to the emission-free channels, and then binned to improve the signal-to-noise ratio. The typical spectral resolution of the reduced data is 1\,km\,s$^{-1}$. The raw spectra are stored in the $T_{\rm{A}}^{\star}$ scale, where $T_{\rm{A}}^{\star}$ is the antenna temperature corrected for the atmospheric attenuation using the chopper-wheel method. The intensity scale is subsequently given in main-beam brightness temperature, $T_{\rm{mb}}$\,=\,$T_{\rm{A}}^{\star}$/$\eta_{\rm{mb}}$, where $\eta_{\rm{mb}}$ is the main-beam efficiency. The adopted main-beam efficiencies are given in Table~\ref{eta_mb} together with frequency, the energy of the upper level, and main beam FWHM for the respective transition. The uncertainty in the absolute intensity scale is estimated to be about $\pm$20\%. Velocities are given with respect to the Local Standard of Rest (LSR).


\subsection{CO line profiles}
\label{ss:coLine}
All new spectra are presented in Appendix~\ref{a:cospectra}, and the integrated intensities and peak main-beam brightness temperatures are given in Tables~\ref{COintensities1}-\ref{COintensities3}. Our previous analysis of CO emission lines from S-stars \citep{ramsetal06} was based on new APEX observations in combination with data from the literature. When examining the previously published data, we found a large scatter in the reported CO($J$\,=\,2$\rightarrow$\,1) intensities and consequently these data were not used in the final analysis. The new CO($J$\,=\,2$\rightarrow$\,1) data is generally stronger than the data discarded in \citet{ramsetal06}. As an example, the new CO($J$\,=\,2\,$\rightarrow$\,1) line intensities obtained at IRAM are, on average, a factor of about three stronger than those reported in \citet{sahaliec95}. 

The observed line profiles are in general of good quality (S/N\,$\sim$\,5 or larger) and can, to a first approximation, be reconciled with those expected from a smooth spherically symmetric outflow. There is no evidence for any "detached-shell" sources in the sample (see e.g. the carbon star sample results of Olofsson et al. 1993\nocite{olofetal93}). A closer inspection shows that some of the stars (e.g., W And, W Aql, DY Gem, and R Gem) have line profiles with weak wings extending beyond the parabolic line profile. This might be indicative of recent mass-loss-rate modulations in these sources or of asymmetric outflows. The six observed lines toward T Cet all show an asymmetry with stronger emission on the red-shifted side, similar to what is observed due to strong self-absorption \citep{huggheal86}. However, the star has a derived mass-loss rate of only 4$\times$10$^{-8}$\,M$_{\odot}$\,yr$^{-1}$ and all lines are found to be optically thin in the model. Consequently, the model does not reproduce the observed line profiles since strong self-absorption requires optically thick lines. Most likely, this is an indication that T Cet has an asymmetric outflow.


\subsection{SiO line profiles}
\label{ss:sioLine}
A total of 26 stars were detected in circumstellar SiO line emission. All new spectra are presented in Appendix~\ref{a:siospectra}, and the integrated intensities and peak main-beam brightness temperatures are given in Tables~\ref{SiOintensities1}-\ref{SiOintensitiesm1}. The observed spectra are generally of good quality (S/N\,$\sim$\,3 or larger). As previosly discussed by \citet{delgetal03}, the SiO line profiles are often found to be narrower than the CO line profiles. However, many of the SiO lines show weak wings extending beyond the main emission feature and the total velocity widths are similar to those of the CO lines. It should be kept in mind that for poorer quality data, the uncertainty in the line width can be several km\,s$^{-1}$.


\subsection{Dust continuum emission}
\label{s:flux}
The SEDs were constructed using J, H, and K band data (2MASS) and IRAS fluxes.


\section{Radiative transfer modelling}
\label{s:COmod}


\subsection{The circumstellar model}
\label{s:circum}
The CSE is assumed to be spherically symmetric and formed by a constant mass-loss rate. It is assumed to be expanding at a constant velocity, derived from fitting the CO line widths, and to have a micro-turbulent velocity distribution with a Doppler width of 1.0\,km\,$\mathrm{s^{-1}}$. In addition, a thermal contribution to the local line width is added, based on the derived kinetic temperature of the gas. The density structure is obtained from the conservation of mass.


\subsection{Dust emission modelling}
\label{s:dustmod}
The dust radiative transfer was solved using the publicly available code DUSTY\footnote{http://www.pa.uky.edu/$\sim$moshe/dusty/}. Amorphous carbon grains \citep{suh00} or amorphous silicate grains \citep{justtiel92} were assumed based on the IRAS low-resolution spectra (LRS) classification according to \citet{volkcohe89}. Three parameters are adjustable when fitting the model SEDs to the observed flux densities; the dust optical depth at 10 $\mu$m, $\tau_{10}$, the dust temperature at the inner radius of the dust envelope, $T_{\rm{d}}(r_{\rm{i}})$, and the stellar temperature, $T_{\star}$. One large grid per dust type was calculated. The optical depth at 10\,$\mu$m ranges from 0.01 to 3.0 in steps of 10 \%, the inner dust temperature is varied from 500 to 1500\,K, and the stellar temperature from 1800 to 2400\,K, both in steps of 100\,K. Once the grid is calculated, the solution can be scaled to the luminosity and distance of any star and a fit to the observed flux densities can be found \citep{ivezelit97}. For simplicity, the dust grains are assumed to be of the same size with a radius of 0.1\,$\mu$m, and have a density of 2\,g\,cm$^{-3}$ (carbon grains) or 3\,g\,cm$^{-3}$ (silicate grains).


\subsection{Physical properties of the gas content in the CSEs}
\label{ss:co}
The adopted method for the radiative transfer analysis of the circumstellar CO radio line emission has been described in detail in previous articles [\cite{schoolof01} for carbon stars, and \cite{olofetal02} for M-type stars] and therefore only a short description is given here. 

The non-LTE radiative transfer code is based on the Monte Carlo method. The code has been benchmarked to high accuracy against a wide variety of molecular-line radiative transfer codes \citep{zandetal02,vdtaketal07}. In the excitation analysis of the CO molecule, 41 rotational levels are included in each of the two lowest vibrational states ($v$=0 and $v$=1). Energy levels and radiative transition probabilities, as well as collisional rate coefficients for collisions between CO and $\rm{H_{2}}$, are taken from \cite{schoetal05}\footnote{http://www.strw.leidenuniv.nl/$\sim$moldata}. When weighting together collisional rate coefficients for CO in collisions with ortho- and para-H$_{2}$ an ortho-to-para ratio of 3 is adopted. Radiation from the central source (assumed to be a blackbody) and the cosmic microwave background is included. When the dust optical depth could be constrained in the dust radiative transfer analysis (see Sect.~\ref{s:res:dust}), i.e., there is a significant amount of themal dust grains present in the wind, the corresponding modification to the radiation field is also included in the molecular excitation analysis. 

The energy balance equation for the gas is solved self-consistently. Cooling is due to line cooling from CO and $\rm{H_{2}}$ and the adiabatic expansion of the gas. The dominant gas-heating mechanism is collisions between the $\rm{H_{2}}$ molecules and the dust grains. Photoelectric heating is also included and it is mostly important in the outer parts of the CSE. The free parameters describing the dust; the dust-to-gas mass-loss-rate ratio, $\Psi$, the average density of an individual dust grain, $\rm{\rho_{g}}$, and the dust grain radius, $a_{\rm{g}}$, are combined in the parameter $h$ [see \citet{schoolof01} for a definition]. The $h$-parameter thus gives a measure of how efficiently the gas is heated due to collisions with dust grains. Following \citet{schoolof01}, we have adopted an average efficiency factor for momentum transfer, which is constant throughout the CSE, $\langle Q_{\rm{rp}} \rangle^{\dagger}$=3\,$\times$\,10$^{-2}$. 

The radial CO abundance distribution is estimated using the model presented in Mamon et al. (1988). 
The initial abundance of CO relative to $\mathrm{H_{2}}$, is assumed to be 6\,$\times$\,10$^{-4}$. The inner radius of the CSE was taken from the SED fitting when the model could be constrained, and is otherwise assumed to be 5\,$r_{\star}$. A change in the inner radius with a factor of two will affect the resulting line intensities by less than 15\% \citep{schoolof01}.

The mass-loss rate and the $h$-parameter are the remaining free parameters in the CO line modelling.


\subsection{SiO line radiative transfer modelling}
\label{ss:siores}
The same non-LTE Monte Carlo radiative transfer code is used for the analysis of the SiO radio line emission, where 41 rotational levels are included in each of the two lowest vibrational states ($v$=0 and $v$=1). Energy levels and radiative transition probabilities, as well as collisional rate coefficients for collisions between SiO and $\rm{H_{2}}$, are taken from \cite{schoetal05}. The physical properties of the CSEs (mass-loss rates and radial gas temperature distributions) are taken from the results of the CO modelling. All other assumptions are the same. Also here, a decrease in the inner radius by a factor of two will result in a small change ($\leq$10\%) of the model line intensities. The only free parameters when fitting the SiO line emission is the SiO abundance at the inner radius, and, when several lines were observed, the outer radius of the SiO emitting region (Sect.~\ref{ss:sioabund}). These two parameters can be used to test chemical models of the inner CSE and photodissociation models for the outer envelope, respectively.


\subsection{The radial SiO abundance distribution}
\label{ss:sioabund}
The radial SiO abundance distribution, i.e., the ratio of the number densities of SiO molecules to H$_{2}$ molecules, $f$\,=\,$n$(SiO)/$n$(H$_{2}$), is assumed to be described by a Gaussian

\begin{equation}
f(r)=f_{0}\ \exp \left(-\left(\frac{r}{r_{\rm{e}}}\right)^{2}\right).
\end{equation}

\noindent
For 18 out of 26 detected stars, too few transitions are observed in order to constrain both the abundance and the size of the emitting region. For these stars the extent of the SiO envelope is assumed to scale with the density of the CSE according to the scaling law found by \citet{delgetal03},

\begin{equation}
\log \ r_{\rm{e}}=19.2+0.48 \ \log \left(\frac{\dot{M}}{v_{\rm{e}}}\right),
\label{re}
\end{equation}

\noindent
where $\dot{M}$ is the mass-loss rate and $v_{\rm{e}}$ the gas expansion velocity of the wind, both found from the CO line modelling. For a discussion about the validity of Eq.~\ref{re} see \citet{delgetal03}, \citet{schoetal06}, and Sect.~\ref{ss:res:sio}.


\section{Results}
\label{s:res}


\subsection{Dust optical depth and radial dust temperature distribution}
\label{s:res:dust}
As known since previous studies of S-stars on the AGB \citep[e.g.,][]{sahaliec95,joriknap98}, the dust content of S-star CSEs is low. The dust radiative transfer analysis performed here show that the dust optical depth can be constrained for only 12 out of 40 sample stars. For the other stars the SED is dominated by the emission from the central star and DUSTY has been used to determine only the stellar effective temperature (in addition to the distance/luminosity of the source). The best-fit model is found by minimizing

\begin{equation}
\chi^{2}=\sum_{\rm{i}=1}^{N} \frac{(F_{\rm{mod,\lambda}}-F_{\rm{obs,\lambda}})^{2}}{\sigma_{\lambda}^{2}},
\end{equation}

\noindent
where $F_{\lambda}$ is the flux density and $\sigma_{\lambda}$ the uncertainty in the measured flux density at wavelength $\lambda$. The summation is done over all $N$ independent observations.
The reduced $\chi^{2}$ for the best-fit model is given by

\begin{equation}
\chi^{2}_{\rm{red}}=\frac{\chi^{2}_{\rm{min}}}{N-p},
\label{chired}
\end{equation}

\noindent
where $p$ (the number of adjustable parameters) is 3 in our case.  

Table~\ref{dust_results1} shows the LRS classification, the assumed dust type, the stellar radius, $r_{\star}$, the inner radius of the CSE, $r_{\rm{i}}$, and the stellar temperature, $T_{\star}$. The inner radius is assumed to be 5\,$r_{\star}$ for all stars where the dust optical depth could not be constrained. Table~\ref{dust_results1} also gives the temperature at the inner radius of the dust shell, $T_{\rm{d}}(r_{\rm{i}})$, the dust optical depth at 10\,$\mu$m, $\tau_{10}$, the dust-to-gas mass-loss-rate ratio, $\Psi$, and the dust mass-loss rate, $\dot{M}_{\rm{d}}$, for the stars where this could be calculated. The dust mass-loss rate is given by

\begin{equation}
\dot{M}_{\rm{d}}=\frac{2\pi}{\chi_{\rm{gr},\nu}} \ v_{d,\infty} \ r_{\rm{i}} \ \tau_{\rm{d}\nu},
\end{equation}

\noindent
where $\chi_{\rm{gr},\nu}$ is the grain cross section per unit mass at a given frequency $\nu$, $v_{d,\infty}$ is the dust expansion velocity, and $\tau_{\rm{d}\nu}$ is the dust optical depth at a given frequency. A $\chi_{\rm{gr},10\mu m}$ of 1.1$\times$10$^{3}$ and 3.5$\times$10$^{3}$ has been used for the carbon stars and M-type stars, respectively. The dust expansion velocity is calculated from the gas expansion velocity derived in the CO line modelling and the drift velocity between the dust and gas particles, $v_{\rm{dr}}$, given by

\begin{equation}
v_{\mathrm{dr}}=\left( \frac{L_{\star} v_{\rm{e}} \langle Q_{\rm{rp}} \rangle^{\dagger}}{\dot{M} c} \right)^{1/2}
\end{equation}

\noindent
\citep{kwok75}.

\begin{table*}[h]
\caption{LRS classification, adopted dust type (amorphous silicate, AMS, or amorphous carbon, AMC) , stellar radius, $r_{\star}$, inner radius of the CSE, $r_{\rm{i}}$, stellar temperature, $T_{\star}$, temperature at the inner boundary, $T_{\rm{d}}(r_{\rm{i}})$, dust optical depth at 10\,$\mu$m, $\tau_{10}$, drift velocity, $v_{\rm{dr}}$, dust mass-loss rate, $\dot{M}_{\rm{d}}$, and dust-to-gas mass-loss-rate ratio, $\Psi$.}
\label{dust_results1}
$$
\begin{array}{p{0.1\linewidth}ccccccccccccccccccc}
\hline
\noalign{\smallskip}
\multicolumn{1}{l}{\rm{Source}} &
\multicolumn{1}{c}{\rm{LRS}} &&
\multicolumn{1}{c}{\rm{Dust}} &&
\multicolumn{1}{c}{r_{\star}} &&
\multicolumn{1}{c}{r_{i}} &&
\multicolumn{1}{c}{T_{\star}} &&
\multicolumn{1}{c}{T_{\rm{d}}(r_{\rm{i}})} &&
\multicolumn{1}{c}{\tau_{10}} &&
\multicolumn{1}{c}{v_{\rm{dr}}} &&
\multicolumn{1}{c}{\dot{M}_{\rm{d}}} &&
\multicolumn{1}{c}{\Psi} \\ 
 &  
\multicolumn{1}{c}{\rm{class}} &&
\multicolumn{1}{c}{\rm{type}} &&
\multicolumn{1}{c}{[10^{13} \ \rm{cm}]} &&
\multicolumn{1}{c}{[10^{14} \ \rm{cm}]} &&
\multicolumn{1}{c}{\rm{[K]}} &&
\multicolumn{1}{c}{\rm{[K]}} && &&
\multicolumn{1}{c}{\rm{[km \ s^{-1}]}} &&
\multicolumn{1}{c}{[10^{-9} \ \rm{M}_{\odot} \ \rm{yr}^{-1}]} &&
\multicolumn{1}{c}{[10^{-3}]} \\
\noalign{\smallskip}
\hline
\noalign{\smallskip}
\object{R And} & \rm{E} && \rm{AMS} && 5.6 && 4.9 && 1800 && \phantom{1}500 && 0.02 && \phantom{1}6.8 && 0.4 && \phantom{1}0.6 \\
\object{W And} & \rm{E} && \rm{AMS} && 3.1 && 1.5 && 2400 && \phantom{1}\cdots && \cdots && \cdots && \cdots && \phantom{1}\cdots  \\
\object{Z Ant} & \rm{E} && \rm{AMS} && 1.6 && 8.2 && 2400 && \phantom{1}\cdots && \cdots && \cdots && \cdots && \phantom{1}\cdots \\
\object{VX Aql} & \cdots && \rm{AMC} && 2.6 && 1.3 && 2400 && \phantom{1}\cdots && \cdots && \cdots && \cdots && \phantom{1}\cdots \\
\object{W Aql} &\rm{E} && \rm{AMS} && 6.0 && 3.7 && 1800 && \phantom{1}600 && 0.10 && \phantom{1}5.7 && 2.4 && \phantom{1}1.1 \\	
\object{AA Cam} & \cdots && \rm{AMC} && 2.6 && 1.3 && 2400 && \phantom{1}\cdots && \cdots && \cdots && \cdots && \phantom{1}\cdots \\
\object{T Cam} & \rm{S} && \rm{AMC} && 3.0 && 1.5 && 2400 && \phantom{1}\cdots && \cdots && \cdots && \cdots && \phantom{1}\cdots \\
\object{S Cas} & \rm{E} && \rm{AMS} && 6.5 && 3.0 && 1800 && \phantom{1}700 && 0.10 && \phantom{1}5.3 && 2.2 && \phantom{1}0.6 \\
\object{V365 Cas} & \rm{F} && \rm{AMC} && 2.6 && 1.3 && 2400 && \phantom{1}\cdots && \cdots && \cdots && \cdots && \phantom{1}\cdots \\
\object{WY Cas} & \rm{E} && \rm{AMS} && 4.0 && 2.0 && 2200 && \phantom{1}\cdots && \cdots && \cdots && \cdots && \phantom{1}\cdots \\	
\object{AM Cen} & \cdots && \rm{AMC} && 2.6 && 1.3 && 2400 &&\phantom{1}\cdots && \cdots && \cdots && \cdots && \phantom{1}\cdots \\
\object{TT Cen} & \cdots && \rm{AMS} && 3.3 && 1.7 && 2400 && \phantom{1}\cdots && \cdots && \cdots && \cdots && \phantom{1}\cdots \\
\object{UY Cen} & \rm{C} && \rm{AMC} && 2.6 && 0.5 && 2400 && 1500 && 0.03 && 15\phantom{.0} && 0.4 && \phantom{1}2.9\\
\object{V386 Cep} & \cdots && \rm{AMS} && 4.6 && 11\phantom{.5} && 1800 && \phantom{1}300 && 0.04 && 14\phantom{.0} && 3.8 && 19\phantom{.2} \\
\object{T Cet} & \cdots && \rm{AMC} && 2.6 && 1.3 && 2400 && \phantom{1}\cdots && \cdots && \cdots && \cdots && \phantom{1}\cdots \\	
\object{AA Cyg} & \rm{S} && \rm{AMC} && 2.6 && 1.3 && 2400 && \phantom{1}\cdots && \cdots && \cdots && \cdots && \phantom{1}\cdots \\
\object{AD Cyg} & \rm{E} && \rm{AMC} && 4.6 && 2.3 && 1800 && \phantom{1}\cdots && \cdots && \cdots && \cdots && \phantom{1}\cdots \\
\object{R Cyg} & \rm{E} && \rm{AMS} && 3.8 && 1.9 && 2200 && \phantom{1}\cdots && \cdots && \cdots && \cdots && \phantom{1}\cdots \\	
\object{$\chi$ Cyg} & \rm{E} && \rm{AMS} && 3.1 && 0.6 && 2400 && 1400 && 0.03 && \phantom{1}9.0 && 0.1 && \phantom{1}0.2 \\
\object{TV Dra}	& \cdots && \rm{AMS} && 2.6 && 1.3 && 2400 && \phantom{1}\cdots && \cdots && \cdots && \cdots && \phantom{1}\cdots \\
\object{DY Gem} & \rm{F} && \rm{AMS} && 2.6 && 1.3 && 2400 && \phantom{1}\cdots && \cdots && \cdots && \cdots && \phantom{1}\cdots \\		
\object{R Gem} & \rm{F} && \rm{AMC} && 3.0 && 1.5 && 2400 && \phantom{1}\cdots && \cdots && \cdots && \cdots && \phantom{1}\cdots \\
\object{ST Her} & \cdots && \rm{AMS} && 3.1 && 3.1 && 2200 && \phantom{1}600 && 0.03  && 13\phantom{.0} && 0.6 && \phantom{1}4.4 \\
\object{RX Lac} & \rm{S} && \rm{AMC} && 2.6 && 1.3 && 2400 && \phantom{1}\cdots && \cdots && \cdots && \cdots && \phantom{1}\cdots \\
\object{GI Lup} & \rm{F} && \rm{AMC} && 2.9 && 1.5 && 2400 && \phantom{1}\cdots && \cdots && \cdots && \cdots && \phantom{1}\cdots \\	
\object{R Lyn} & \rm{F} && \rm{AMC} && 3.0 && 1.5 && 2400 && \phantom{1}\cdots && \cdots && \cdots && \cdots && \phantom{1}\cdots \\
\object{Y Lyn} & \cdots && \rm{AMS} && 1.9 && 1.0 && 2400 && \phantom{1}\cdots && \cdots && \cdots && \cdots && \phantom{1}\cdots \\	
\object{S Lyr} & \cdots && \rm{AMS} && 5.7 && 5.1 && 1800 && \phantom{1}500 && 0.09 && \phantom{1}5.0 && 2.4 && \phantom{1}1.2 \\
\object{FU Mon} & \rm{S} && \rm{AMC} && 2.6 && 1.3 && 2400 && \phantom{1}\cdots && \cdots && \cdots && \cdots && \phantom{1}\cdots \\
\object{RZ Peg} & \rm{C} && \rm{AMC} && 2.6 && 1.3 && 2400 && \phantom{1}\cdots && \cdots && \cdots && \cdots && \phantom{1}\cdots \\		
\object{RT Sco} & \rm{E} && \rm{AMS} && 3.3 && 1.7 && 2400 && \phantom{1}\cdots && \cdots && \cdots && \cdots && \phantom{1}\cdots \\ 
\object{ST Sco} & \rm{E} && \rm{AMS} && 2.6 && 1.3 && 2400 && \phantom{1}\cdots && \cdots && \cdots && \cdots && \phantom{1}\cdots \\
\object{RZ Sgr} & \rm{F} && \rm{AMC} && 2.6 && 1.3 && 2400 && \phantom{1}\cdots && \cdots && \cdots && \cdots && \phantom{1}\cdots \\		
\object{ST Sgr} & \rm{E} && \rm{AMS} && 3.1 && 1.6 && 2400 && \phantom{1}\cdots && \cdots && \cdots && \cdots && \phantom{1}\cdots \\
\object{T Sgr} & \rm{F} && \rm{AMC} && 3.1 && 1.6 && 2400 && \phantom{1}\cdots && \cdots && \cdots && \cdots && \phantom{1}\cdots \\
\object{EP Vul} & \rm{F} && \rm{AMC} && 2.6 && 1.3 && 2400 && \phantom{1}\cdots && \cdots && \cdots && \cdots && \phantom{1}\cdots \\
\object{DK Vul} & \rm{F} && \rm{AMC} && 2.6 && 1.3 && 2400 && \phantom{1}\cdots && \cdots && \cdots && \cdots && \phantom{1}\cdots \\
\object{AFGL 2425} & \cdots && \rm{AMS} && 4.6 && 2.9 && 1800 && \phantom{1}600 && 0.10 && \phantom{1}8.4 && 1.4 && \phantom{1}4.8 \\
\object{CSS2 41} & \cdots && \rm{AMS} && 4.6 && 4.1 && 1800 && \phantom{1}500 && 0.10 && \phantom{1}8.5 && 3.0 && \phantom{1}5.2 \\
\object{IRC--10401}	& \cdots && \rm{AMS} && 4.6 && 4.1 && 1800 && \phantom{1}500 && 0.07 && 11\phantom{.0} && 2.3 && \phantom{1}6.6 \\
\noalign{\smallskip}
\hline
\end{array}
$$
\end{table*}
%


%
\begin{table}[h]
\caption{Mass-loss rates, $h$-parameters, and terminal gas expansion velocities, $v_{\rm{e}}$, derived from fitting the radiative transfer model to the CO line observations.}
\label{Mdot}
$$
\begin{array}{p{0.2\linewidth}ccccccrccc}
\hline
\noalign{\smallskip}
\multicolumn{1}{l}{\rm{Source}} &
\multicolumn{1}{c}{\dot{M}} &&
\multicolumn{1}{c}{h} &&
\multicolumn{1}{c}{v_{\rm{e}}} &&
\multicolumn{1}{c}{v_{\rm{LSR}}} &&
\multicolumn{1}{c}{\chi^{2}_{\rm{red}}} &
\multicolumn{1}{c}{N} \\ 
 &
\multicolumn{1}{c}{[10^{-7} \ \rm{M}_{\odot} \ \rm{yr}^{-1}]} &&
 &&
\multicolumn{1}{c}{[\rm{km \ s}^{-1}]} &&
\multicolumn{1}{c}{[\rm{km \ s}^{-1}]} & \\
\noalign{\smallskip}
\hline
\noalign{\smallskip}
\object{R And} & \phantom{1}6.6 && 0.4\phantom{:} && \phantom{1}8.3 && -16.5 && \phantom{1}1.3 & 5 \\
\object{W And} & \phantom{1}1.7 && 0.5\phantom{:} && \phantom{1}6.0 && -35.0 && \phantom{1}1.1 & 5 \\
\object{Z Ant} & \phantom{1}0.9 && 0.2: && \phantom{1}6.0 && -16.0 && \phantom{1}0.4 & 2 \\
\object{VX Aql} & \phantom{1}1.0 && 0.2: && \phantom{1}7.0 && 6.7 && \cdots & 1 \\
\object{W Aql} & 22\phantom{.0} && 1.0\phantom{:} && 17.2 && -26.0 && \phantom{1}0.6 & 5 \\	
\object{AA Cam} & \phantom{1}0.4 && 0.2: && \phantom{1}3.4 && -46.8 && \phantom{1}2.8 & 2 \\
\object{T Cam} & \phantom{1}1.0 && 0.2: && \phantom{1}3.8 && -11.7 && \phantom{1}0.1 & 2 \\
\object{S Cas} & 35\phantom{.0} && 0.2\phantom{:} && 20.5 && -31.0 && \phantom{1}3.6 & 4 \\		
\object{V365 Cas} & \phantom{1}0.3 && 0.2: && \phantom{1}6.2 && -2.1 && \cdots & 1 \\
\object{WY Cas} & 11\phantom{.0} && 0.5\phantom{:} && 13.5 && 7.0 && \phantom{1}6.6 & 3 \\		
\object{AM Cen} & \phantom{1}1.5 && 0.2: && \phantom{1}3.2 && -27.0 && \phantom{1}2.0 & 2 \\
\object{TT Cen} & 25\phantom{.0} && 0.5: && 20.0 && 4.0 && \phantom{1}7.4 & 2 \\
\object{UY Cen} & \phantom{1}1.3 && 0.2: && 12.0 && -26.0 && \phantom{1}0.2 & 2 \\
\object{V386 Cep} & \phantom{1}2.0 && 0.2: && 16.0 && -51.8 && \cdots & 1 \\
\object{T Cet} & \phantom{1}0.4 && 0.2\phantom{:} && \phantom{1}5.5 && 22.0 && \phantom{1}1.2 & 6 \\	
\object{AA Cyg} & \phantom{1}2.9 && 0.2\phantom{:} && \phantom{1}4.5 && 27.5 && \phantom{1}2.1 & 3 \\
\object{AD Cyg} & \phantom{1}2.1 && 0.2: && \phantom{1}8.0 && 21.0 && \phantom{1}0.1 & 2 \\
\object{R Cyg} & \phantom{1}6.3 && 0.5\phantom{:} && \phantom{1}9.0 && -18.5 && \phantom{1}5.0 & 5 \\		
\object{$\chi$ Cyg} & \phantom{1}3.8 && 0.2\phantom{:} && \phantom{1}8.5 && 9.0 && \phantom{1}2.0 & 6 \\
\object{TV Dra}	& \phantom{1}0.5 && 0.2\phantom{:} && \phantom{1}4.7 && 22.0 && \phantom{1}7.8 & 3 \\
\object{DY Gem} & \phantom{1}7.0 && 0.2\phantom{:} && \phantom{1}8.0 && -16.7 && \phantom{1}2.0 & 3 \\
\object{R Gem} & \phantom{1}4.4 && 0.8\phantom{:} && \phantom{1}4.5 && -60.0 && \phantom{1}3.7 & 4 \\
\object{ST Her} & \phantom{1}1.3 && 0.2\phantom{:} && \phantom{1}8.5 && -4.5 && \phantom{1}1.4 & 4 \\
\object{RX Lac} & \phantom{1}0.8 && 0.2: && \phantom{1}6.5 && -15.4 && \cdots & 1 \\
\object{GI Lup} & \phantom{1}5.5 && 0.5: && 10.0 && 6.0 && \cdots & 1 \\		
\object{R Lyn} & \phantom{1}3.3 && 0.2\phantom{:} && \phantom{1}7.5 && 16.0 && \phantom{1}0.5 & 5 \\
\object{Y Lyn} & \phantom{1}2.3 && 1.0\phantom{:} && \phantom{1}7.5 && -0.5 && \phantom{1}3.9 & 3 \\	
\object{S Lyr} & 20\phantom{.0} && 0.5: && 13.0 && 49.0 && \phantom{1}7.6 & 2 \\	
\object{FU Mon} & \phantom{1}2.7 && 0.2\phantom{:} && \phantom{1}2.8 && -42.0 && \phantom{1}0.6 & 3 \\
\object{RZ Peg} & \phantom{1}4.6 && 0.2: && 12.6 && -23.4 && \cdots & 1 \\
\object{RT Sco} & \phantom{1}4.5 && 0.5: && 11.0 && -47.0 &&  \phantom{1}3.2 & 2 \\
\object{ST Sco} & \phantom{1}1.5 && 0.2\phantom{:} && \phantom{1}5.5 && -4.5 && \phantom{1}2.0 & 3 \\
\object{RZ Sgr} & 30\phantom{.0} && 0.6\phantom{:} && \phantom{1}9.0 && -31.0 && \phantom{1}1.4 & 3 \\		
\object{ST Sgr} & \phantom{1}2.0 && 0.5: && \phantom{1}6.0 && 55.7 && \phantom{1}0.1 & 2 \\
\object{T Sgr} & \phantom{1}1.4 && 0.2\phantom{:} && \phantom{1}7.5 && 9.0 && \phantom{1}3.8 & 3 \\	
\object{EP Vul} & \phantom{1}2.0 && 0.2\phantom{:} && \phantom{1}5.7 && 0.0 && \phantom{1}8.9 & 3 \\
\object{DK Vul} & \phantom{1}2.5 && 0.6\phantom{:} && \phantom{1}4.5 && -15.0 && \phantom{1}5.7 & 4 \\
\object{AFGL 2425} & \phantom{1}3.0 && 0.2\phantom{:} && \phantom{1}8.7 && 57.0 && \phantom{1}0.2 & 3 \\
\object{CSS2 41} & \phantom{1}5.8 && 0.2\phantom{:} && 17.0 && 21.5 && \phantom{1}2.4 & 3 \\	
\object{IRC--10401}	& \phantom{1}3.5 && 0.2: && 17.0 && 19.0 && 13.4 & 2 \\
\noalign{\smallskip}
\hline
\end{array}
$$
\end{table}
\begin{figure*}[t]
\raggedright 
{\includegraphics[width=18cm]{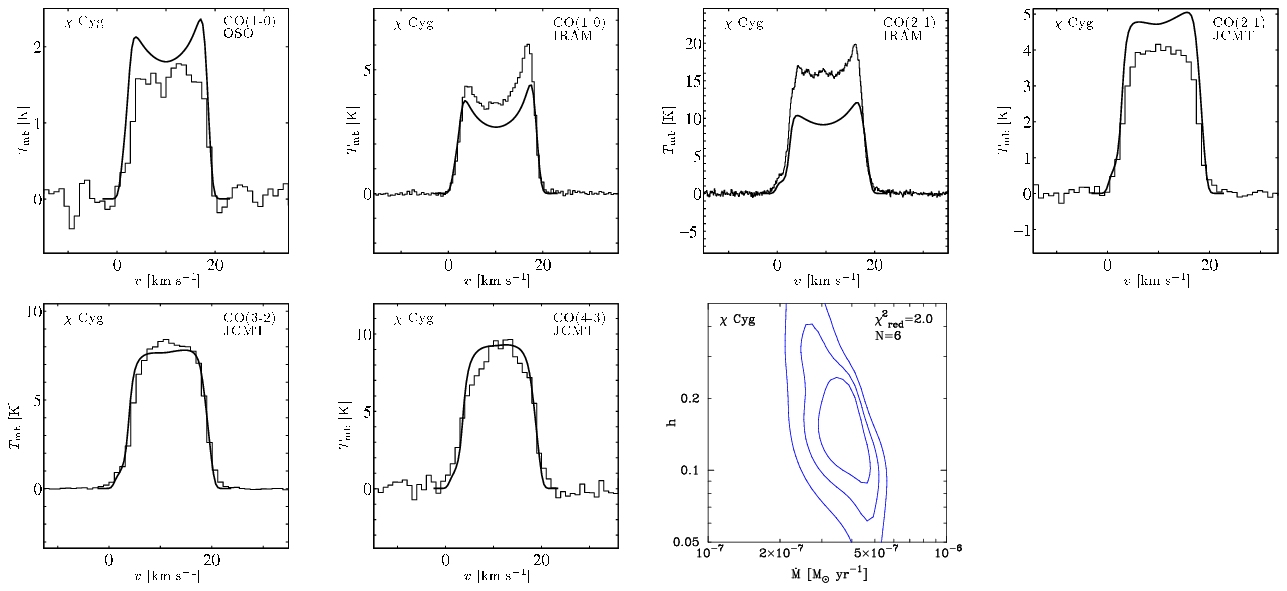}}
\caption{The observed CO line profiles toward $\chi$ Cyg (histogram) overlayed with the results from the radiative transfer model (solid line) for the best-fit model with $\chi^{2}_{\rm{red}}$=2.0. The $\chi^{2}$-map of the full analysis of $\chi$ Cyg is also shown. LSR velocity scales are used.}
\label{fincomodel}
\end{figure*}

\subsection{Mass-loss rates, radial gas temperature distribution and kinematics}
\label{ss:res:mdot}
All results from the CO radiative transfer analysis, mass-loss rates, $h$-parameters, gas expansion velocities, stellar velocities, $v_{\rm{LSR}}$, $\chi^{2}_{\rm{red}}$ for the best-fit models, and the number of observational constraints, $N$, are given in Table~\ref{Mdot}.

For slightly more than half of the stars, enough data ($N\geq3$) is available to constrain both $h$ and $\dot{M}$. Models are calculated for a large number of mass-loss rates and $h$-parameters and the best-fit model is found by minimizing

\begin{equation}
\chi^{2}=\sum_{\rm{i}=1}^{N} \frac{(I_{\rm{mod,i}}-I_{\rm{obs,i}})^{2}}{\sigma^{2}_{\rm{i}}},
\label{chi2I}
\end{equation}

\noindent
where $I_{\rm{mod}}$ and $I_{\rm{obs}}$ are the integrated intensities of the modelled and observed lines, respectively. $\sigma$ is in most cases dominated by the calibration uncertainties assumed to be 20\%. For observations with poorer S/N, $\sigma$ is set higher according to the quality of the data (although never above 30\%). The reduced $\chi^{2}$ is given by Eq.~\ref{chired}, where $p=2$.

For the remaining stars, we have assumed the value of $h$ \,Êdepending on the stellar luminosity; $h$=0.2 for $L_{\star} <5000 \,L_{\odot}$ and $h$=0.5 for $L_{\star} \geq5000 \,L_{\odot}$ [in accordance with \citet{schoolof01} and \citet{olofetal02}]. This is indicated by a colon in Table~\ref{Mdot} and the reduced $\chi^{2}$ is given by Eq.~\ref{chired} with $p=1$.

Generally, the mass-loss rates are well constrained, and the observed CO lines are well-reproduced by the model. For many of the stars, it is not possible to find an upper limit to the $h$-parameter. This is not surprising, since the excitation in these low-mass-loss-rate stars is dominated by the radiation from the central star and is not very sensitive to the temperature structure of the gas and the adopted dust parameters. For some stars the $\chi^{2}_{\rm{red}}$ of the best-fit model is rather large. The derived $\chi^2$ is very sensitive to the line ratios, and any calibration errors in one line (larger than our estimated uncertainties) might result in a large $\chi^{2}_{\rm{red}}$ (see Fig.~\ref{fincomodel}, CO($J$\,=\,2$\rightarrow$\,1)). Therefore it is difficult to determine whether inadequacies in the model or calibration errors are responsible for the poor fit in some cases. We find no systematic trends with particular lines or telescopes. Figure~\ref{fincomodel} shows the best-fit model from the radiative transfer analysis overlayed on the observed line profiles for $\chi$ Cyg. The $\chi^{2}$-map is also shown with the innermost contour representing the 1$\sigma$-level. This fit is good, but not perfect, e.g., in the details of the line shapes, and the main reason for this is not easily identified.

Compared to the mass-loss rates found in \citet{ramsetal06}, the larger quantity of data makes the new mass-loss-rate estimates more reliable. In particular, it is possible to determine the $h$-parameter for a larger number of stars. A comparison shows that the inclusion of the new data in the analysis has changed the estimated mass-loss rates {\bf by} less than a factor of two. For seven stars the change is slightly larger. There is no systematic change upwards or downwards.


\subsection{SiO abundances and radial distribution}
\label{ss:res:sio}

\begin{figure}[h]
\raggedright 
{\includegraphics[width=\columnwidth]{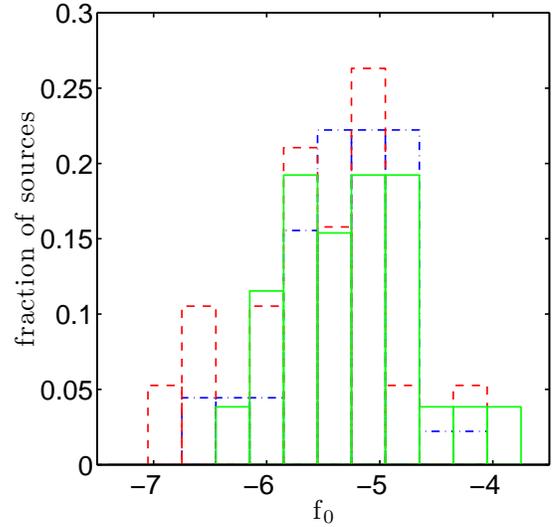}}
\caption{Circumstellar SiO fractional abundances ($f_{0}$) for the S-type (solid, green line; 26 stars), M-type (dashed-dotted, blue line; 45 stars; \citet{delgetal03}), and carbon star (dashed, red line; 19 stars; \citep{schoetal06}) samples.}
\label{plot_sio2}
\end{figure}

All results from the SiO radiative transfer analysis, SiO fractional abundances, $f_{0}$, the size of the SiO emitting regions (as derived from Eq.~\ref{re}), $r_{\rm{e}}$, SiO gas expansion velocities, $v_{\rm{e}}(\rm{SiO})$, $\chi^{2}_{\rm{red}}$ for the best-fit models, and the number of observational constraints, $N$, are given in Table~\ref{abundance}. Again, the best-fit model is found by minimizing Eq.~\ref{chi2I} and $\chi^{2}_{\rm{red}}$ is defined by Eq.~\ref{chired}, where $p$\,=\,1 when $r_{\rm{e}}$ is given by Eq.~\ref{re}. The derived abundances range from $4\times 10^{-7}$ to $1.4\times 10^{-4}$. The median value is $6\times 10^{-6}$. Figure~\ref{plot_sio2} shows a histogram of the circumstellar SiO abundances for the S-type stars compared to those found for M-type stars \citep{delgetal03} and carbon stars \citep{schoetal06}. The spread of the SiO abundance distribution for the S-type stars [as measured by the ratio between the 90th percentile and the 10th percentile] is about a factor of 25.

The SiO gas expansion velocity is found to be, on average, 20\% smaller than the CO gas expansion velocity (for 19 out of 26 stars). \citet{delgetal03} discuss the possibility that the SiO emission probes the very outer parts of the gas acceleration zone and that this would explain the discrepancy in the line width. SiO typically probes regions closer to the star by about a factor of 5--10 compared to CO. A trend of narrowing of the SiO lines with higher frequency might then also be expected, since rotational transitions involving higher energy levels (Table~\ref{eta_mb}) tend to probe warmer and denser regions closer to the star. In addition, the line width can be reduced due to strong self-absorption on the blue-shifted side of the SiO lines that are optically thick. We do not find a trend with narrowing of higher frequency lines nor with the mass-loss rate [as would be expected if self-absorption was the predominant effect] and can therefore not firmly conclude that the SiO emission probes the acceleration zone.

\begin{figure*}[t]
\raggedright 
\sidecaption
{\includegraphics[width=12cm]{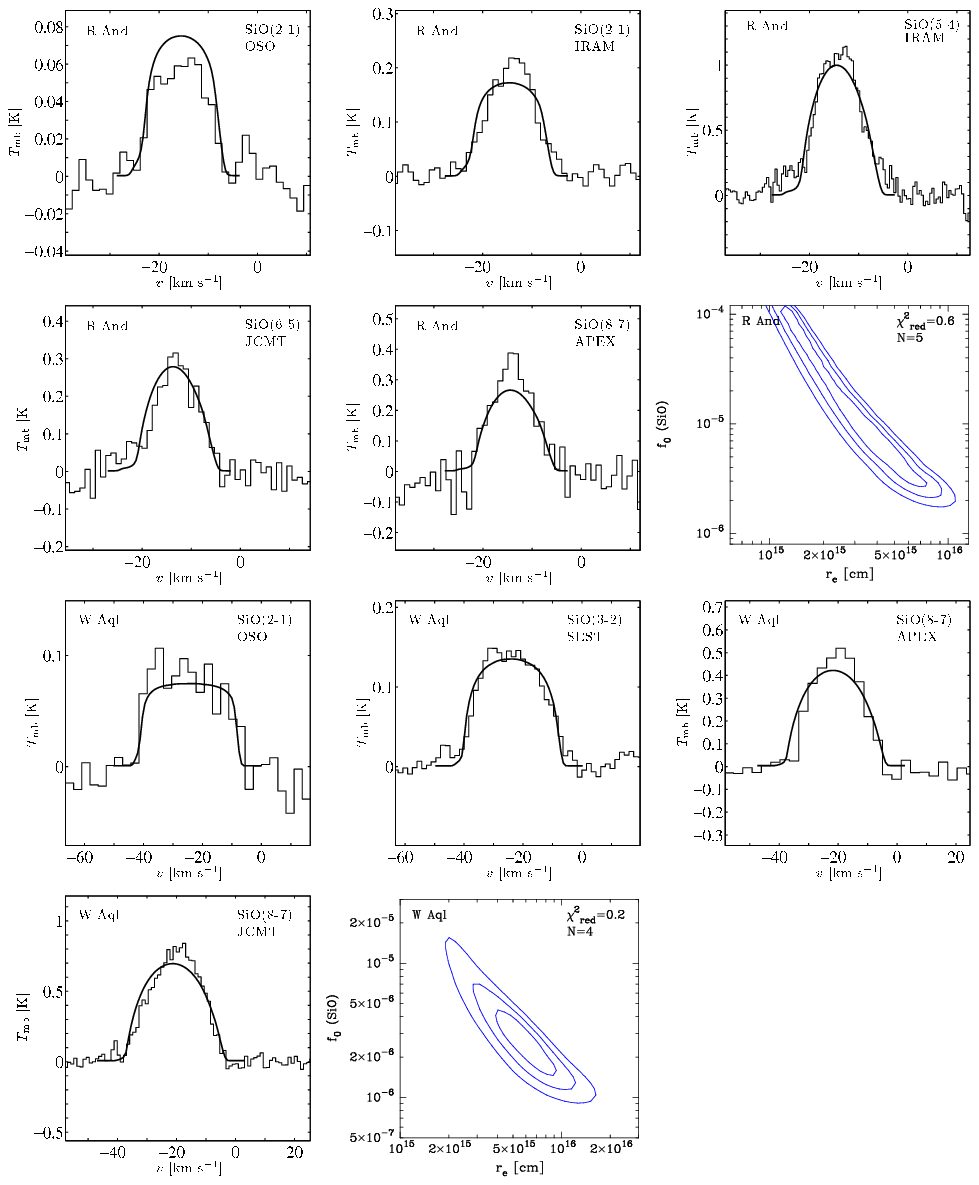}}
\caption{The observed SiO line profiles toward R And and W Aql (histogram) overlayed with the results from the radiative transfer model (solid line) for the best-fit models with $\chi^{2}_{\rm{red}}$=0.6 and 0.2, respectively. LSR velocity scales are used. The $\chi^{2}$-maps of the full analysis for the two stars are also shown.}
\label{finsiomodel}
\end{figure*}

Table~\ref{abund_best} shows the results (with 1$\sigma$-errors) of the best-fit models when $r_{\rm{e}}$ is left as a free parameter. In this case, the modelling generally gives a good fit to the observed SiO lines (see Fig.~\ref{finsiomodel}) with $\chi^{2}_{\rm{red}}$ on the order of unity. From the $\chi^{2}$- analysis we find that the derived SiO abundances are generally determined within a factor of $\approx$\,4 (except for R And, Fig.~\ref{finsiomodel}). S Cas has a large $\chi^{2}_{\rm{red}}$-value and the same can be observed for the result from the CO model (Table~\ref{Mdot}). This might indicate that S Cas is not well reproduced by a spherically symmetric wind model. The same can be noted for the SiO model of $\chi$ Cyg. However, both models are based on rather noisy SiO($J$\,=\,2\,$\rightarrow$1) lines from OSO, and the line ratios between this line and the other available lines are not reproduced. Thus, we can not decide whether the bad fit is due to the model being incorrect for these stars, or observational uncertainties. For T Cet, only the SiO($J$\,=\,8$\rightarrow$7) line was observed (at APEX). A fit to the width of the line gives $v_{\rm{e}}$(SiO)\,=\,16.0 km s$^{-1}$. Since only one line is available and the data is rather noisy (see Fig~\ref{rand_sio1}), we have chosen to model the star with the expansion velocity derived from the CO model ($N$=6), $v_{\rm{e}}$=5.5\,km\,s$^{-1}$. For IRC--10401 enough lines are available to try to fit the size of the SiO envelope and the abundance simultaneously. However, we are not able to constrain $r_{\rm{e}}$ for this object and have chosen to use $r_{\rm{e}}$ from Eq.~\ref{re}.

\begin{table}[h]
\caption{Circumstellar SiO fractional abundance, $f_{0}$, radius of the SiO line-emitting region from Eq.~\ref{re}, $r_{\rm{e}}$, and gas expansion velocities derived from fitting the SiO line radiative transfer model to the observations, $v_{\rm{e}}(\rm{SiO})$. The reduced $\chi^{2}$ for the best-fit model and the number of observational constraints, $N$, are also given.}
\label{abundance}
$$
\begin{array}{p{0.2\linewidth}cccccccc}
\hline
\noalign{\smallskip}
\multicolumn{1}{l}{\rm{Source}} &
\multicolumn{1}{c}{f_{0}} &&
\multicolumn{1}{c}{r_{\rm{e}}} &&
\multicolumn{1}{c}{v_{\rm{e}}(\rm{SiO})} &&
\multicolumn{1}{c}{\chi^{2}_{\rm{red}}} &
\multicolumn{1}{c}{N} \\ 
 &
\multicolumn{1}{c}{[10^{-6}]} &&
\multicolumn{1}{c}{[10^{15} \rm{cm}]} &&
\multicolumn{1}{c}{[\rm{km \ s}^{-1}]} && 
 & \\
\noalign{\smallskip}
\hline
\noalign{\smallskip}
\object{R And} & \phantom{11}4.0 && \phantom{1}6.2 && \phantom{1}8.3\phantom{:} && \phantom{1}1.0 & 5 \\
\object{W And} & \phantom{1}12\phantom{.0} && \phantom{1}3.8 && \phantom{1}4.8\phantom{:} && \phantom{1}0.8 & 2 \\
\object{W Aql} & \phantom{11}2.0 && \phantom{1}7.8 && 17.0\phantom{:} && \phantom{1}0.6 & 4 \\	
\object{S Cas} & \phantom{11}6.0 && \phantom{1}8.9 && 20.5\phantom{:} && \phantom{1}4.2 & 3 \\		
\object{WY Cas} & \phantom{11}1.5 && \phantom{1}6.3 && 10.0\phantom{:} && \phantom{1}2.2 & 3 \\		
\object{V386 Cep} & 140\phantom{.0} && \phantom{1}2.5 && 13.5\phantom{:} && \phantom{1}0.3 & 3 \\
\object{T Cet} & \phantom{1}12\phantom{.0} && \phantom{1}2.0 && \phantom{1}5.5: && \phantom{1}\cdots & 1 \\	
\object{AA Cyg} & \phantom{11}1.9 && \phantom{1}5.6 && \phantom{1}4.0\phantom{:} && \phantom{1}3.7 & 2 \\
\object{R Cyg} & \phantom{11}4.0 && \phantom{1}5.8 && \phantom{1}9.0\phantom{:} && \phantom{1}1.3 & 4 \\	
\object{$\chi$ Cyg} & \phantom{1}10\phantom{.0} && \phantom{1}4.7 && \phantom{1}7.0\phantom{:} && \phantom{1}4.7 & 4 \\
\object{TV Dra}	& \phantom{1}26\phantom{.0} && \phantom{1}2.4 && \phantom{1}4.5\phantom{:} && \phantom{1}5.7 & 2 \\
\object{DY Gem} & \phantom{11}1.8 && \phantom{1}6.5 && \phantom{1}6.5\phantom{:} && \phantom{1}0.5 & 2 \\ 
\object{R Gem} & \phantom{11}1.4 && \phantom{1}6.8 && \phantom{1}4.0\phantom{:} && \phantom{1}2.1 & 3 \\ 
\object{RX Lac} & \phantom{11}3.0 && \phantom{1}2.5 && \phantom{1}3.0\phantom{:} && \phantom{1}2.0 & 3 \\
\object{Y Lyn} & \phantom{11}1.0 && \phantom{1}3.9 && \phantom{1}5.0\phantom{:} && \phantom{1}0.5 & 3 \\  	
\object{S Lyr} & \phantom{1}10\phantom{.0} && \phantom{1}8.5 && 11.5\phantom{:} && \phantom{1}3.0 & 2 \\	
\object{RT Sco} & \phantom{11}1.5 && \phantom{1}4.5 && \phantom{1}9.5\phantom{:} && \phantom{1}0.4 & 2 \\ 
\object{ST Sco} & \phantom{11}1.0 && \phantom{1}3.7 && \phantom{1}5.5\phantom{:} && \phantom{1}7.9 & 2 \\ 
\object{RZ Sgr} & \phantom{11}0.4 && 12\phantom{.0} && \phantom{1}7.0\phantom{:} && \phantom{1}0.2 & 2 \\  
\object{ST Sgr} & \phantom{1}10\phantom{.0} && \phantom{1}4.1 && \phantom{1}6.0\phantom{:} && \phantom{1}\cdots & 1 \\ 
\object{T Sgr} & \phantom{1}22\phantom{.0} && \phantom{1}3.1 && \phantom{1}5.0\phantom{:} && \phantom{1}\cdots & 1 \\  	
\object{EP Vul} & \phantom{1}10\phantom{.0} && \phantom{1}4.2 && \phantom{1}5.0\phantom{:} && \phantom{1}1.3 & 3 \\  
\object{DK Vul} & \phantom{11}3.8 && \phantom{1}5.9 && \phantom{1}3.5\phantom{:} && \phantom{1}0.4 & 2 \\
\object{AFGL 2425} & \phantom{1}16\phantom{.0} && \phantom{1}4.2 && \phantom{1}8.7\phantom{:} && \phantom{1}3.3 & 3 \\
\object{CSS2 41} & \phantom{1}68\phantom{.0} && \phantom{1}4.1 && 16.0\phantom{:} && \phantom{1}0.3 & 2 \\  
\object{IRC--10401}	& \phantom{1}14\phantom{.0} && \phantom{1}3.2 && 17.0\phantom{:} && 22.8 & 3 \\
\noalign{\smallskip}
\hline
\end{array}
$$
\end{table}
\begin{table}[h]
\caption{Circumstellar SiO fractional abundance, $f_{0}$, and the radius of the SiO line-emitting region, $r_{\rm{e}}$, with 1$\sigma$-errors, from the best-fit model. The reduced $\chi^{2}$ for the best-fit model and the number of observational constraints, $N$, are also given.}
\label{abund_best}
$$
\begin{array}{p{0.2\linewidth}cccccc}
\hline
\noalign{\smallskip}
\multicolumn{1}{l}{\rm{Source}} &
\multicolumn{1}{c}{f_{0}} &&
\multicolumn{1}{c}{r_{\rm{e}}} &&
\multicolumn{1}{c}{\chi^{2}_{\rm{red}}} &
\multicolumn{1}{c}{N} \\ 
 & && &&
 & \\
\noalign{\smallskip}
\hline
\noalign{\smallskip}
\object{R And} & 5.1\pm4.9\times10^{-5} && 4.0\pm3.0\times10^{15} && 0.6 & 5 \\
\object{W Aql} & 3.0\pm1.0\times10^{-6} && 6.5\pm2.5\times10^{15} && 0.2 & 4 \\	
\object{S Cas} & 9.0\pm6.0\times10^{-6} && 9.0\pm4.0\times10^{15} && 7.2 & 3 \\		
\object{WY Cas} & 1.2\pm0.3\times10^{-6} && 1.3\pm0.7\times10^{16} && 2.2 & 3 \\		
\object{V386 Cep} & 7.5\pm2.5\times10^{-5} && 3.0\pm2.0\times10^{15} && 0.7 & 3 \\
\object{R Cyg} & 3.0\pm1.0\times10^{-6} && 8.5\pm3.5\times10^{15} && 1.2 & 4 \\		
\object{$\chi$ Cyg} & 1.3\pm0.7\times10^{-5} && 5.0\pm3.0\times10^{15} && 5.2 & 4 \\
\object{R Gem} & 1.3\pm1.1\times10^{-5} && 3.0\pm2.0\times10^{15} && 1.1 & 3 \\ 
\object{RX Lac} & 1.8\pm0.8\times10^{-6} && 3.6\pm2.4\times10^{15} && 1.0 & 3 \\
\object{Y Lyn} & 1.9\pm1.1\times10^{-6} && 3.5\pm1.5\times10^{15} && 0.9 & 3 \\ 
\object{EP Vul} & 2.3\pm1.7\times10^{-5} && 3.1\pm2.0\times10^{15} && 0.7 & 3 \\  
\object{AFGL 2425} & 6.5\pm2.5\times10^{-6} && 4.3\pm3.5\times10^{16} && 1.2 & 3 \\ 	
\noalign{\smallskip}
\hline
\end{array}
$$
\end{table}

\begin{figure}[h]
\raggedright 
{\includegraphics[width=\columnwidth]{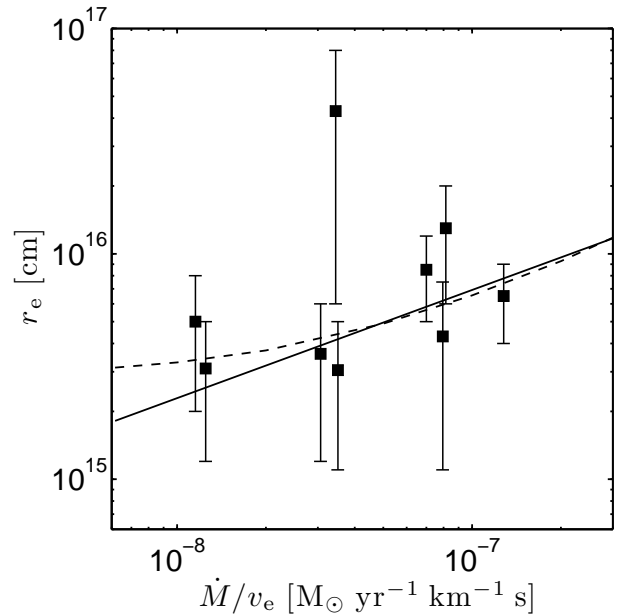}}
\caption{Circumstellar SiO envelope sizes ($r_{\rm{e}}$) estimated from the best-fit models plotted against the wind density ($\dot{M}/v_{\rm{e}}$). The solid line is derived from Eq.~\ref{re}, which is based on a fit to SiO envelope estimates for a sample of M-type AGB stars \citep{delgetal03}. The dashed line shows the prediction using a simple photochemical model (see Sect.~\ref{ss:dis:sio}).}
\label{size}
\end{figure}

To be able to compare our results to what has previously been found for M-type stars \citep{delgetal03} and carbon stars \citep{schoetal06} we have decided to derive all SiO abundances using Eq.~\ref{re} for the size of the SiO emitting region in order to be consistent. Figure~\ref{size} shows a comparison between the radii derived when $r_{\rm{e}}$ is left as a free parameter and those using Eq.~\ref{re}. The error bars correspond to the 1$\sigma$ limits. Only results from models with a $\chi^{2}_{\rm{red}}<2.5$ and where the SiO($J$\,=\,8$\rightarrow$7) line was available are shown in Fig.~\ref{size}. We conclude that our results are consistent with Eq.~\ref{re} and that the size of the SiO emitting region does not differ significantly between the different chemical types. This is in agreement with the results of \citet{schoetal06}. 

%


\section{Discussion}
\label{s:discuss}


\subsection{Mass loss of S-type AGB stars}
\label{ss:dis:mdot}

\begin{figure*}[t]
\raggedright 
{\includegraphics[width=18cm]{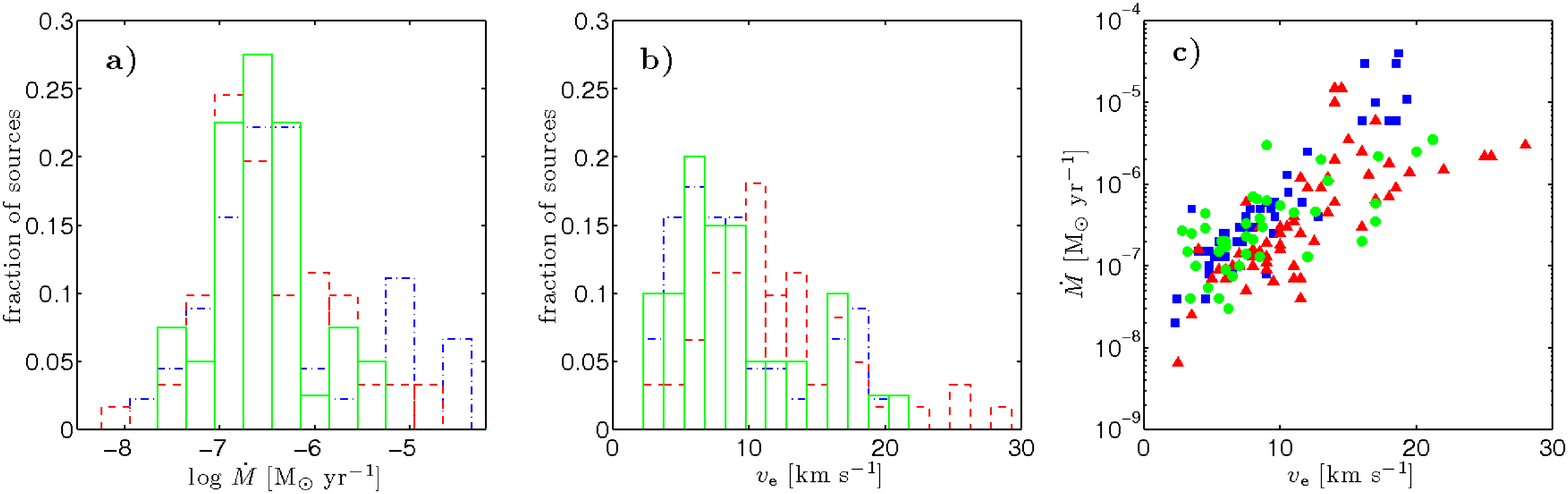}}
\caption{{\bf a)} Mass-loss-rate distributions for S-type stars (solid, green line; 40 stars), M-type stars (dashed-dotted, blue line), and carbon AGB stars (dashed, red line) samples. {\bf b)} Gas expansion velocity distributions derived from fitting the CO line widths for the S-type (solid, green line), M-type (dotted, blue line) and carbon AGB star (dashed, red line) samples. {\bf c)} Mass-loss rates plotted against the gas expansion velocities for S-type stars (green dots), M-type stars (blue squares), and carbon AGB stars (red triangles) samples.} 
\label{plot_co}
\end{figure*}

Figure~\ref{plot_co}a shows our mass-loss-rate results for the S-type AGB star sample compared with previous results for carbon stars \citep{schoolof01} and M-type AGB stars \citep{olofetal02,delgetal03}. We confirm here the result of \citet{ramsetal06} that the mass-loss-rate distributions for the three different samples are very similar. The median value for the S-type stars using the new data is 2.7$\times$10$^{-7}$\,M$_{\odot}$\,yr$^{-1}$, and the M-type and carbon stars both have a median of 3.0$\times$10$^{-7}$\,M$_{\odot}$\,yr$^{-1}$. There are possibly fewer S-type stars with high mass-loss rates and due to our sample selection criteria (see Sect.~\ref{s:sample}) S-type stars with no or very little mass loss will be missed. The derived mass-loss rates depend on the adopted CO fractional abundance, which differs for the three chemical types in accordance with stellar atmosphere models. Figure~\ref{plot_co}b shows the gas expansion velocity distribution derived from the CO line widths for the three samples. The median gas expansion velocity for the S-type stars is 8\,km\,s$^{-1}$, and 7.5 and 11\,km\,s$^{-1}$ for the M-type and carbon stars, respectively, indicating that the carbon stars have CSEs with higher gas expansion velocities. Finally, Fig.~\ref{plot_co}c shows the relation between the mass-loss rates and the expansion velocities for the three samples. We find no apparent difference depending on chemistry (apart from the higher gas expansion velocities in the carbon star sample) and suggest that this points to a mass loss that is driven by the same mechanism(s) in the S-type, M-type and carbon AGB stars. 

\begin{figure*}[t]
\raggedright 
{\includegraphics[width=18cm]{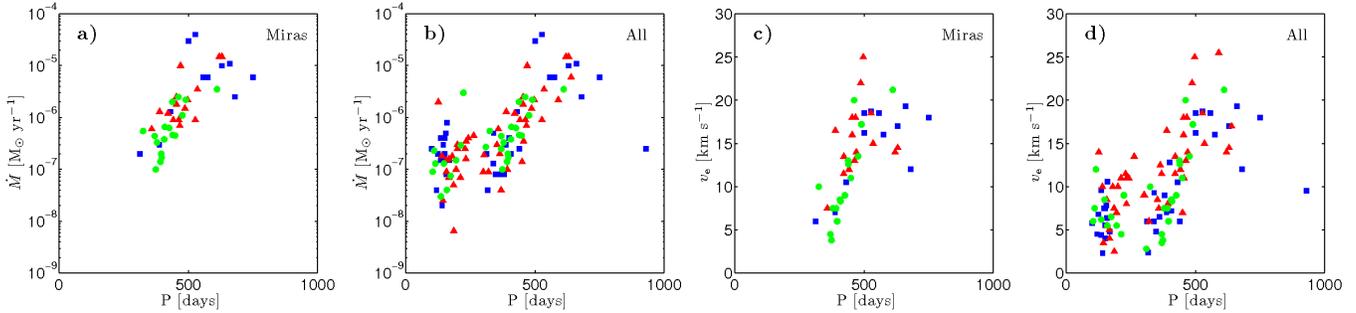}}
\caption{Mass-loss rates plotted against the periods of the Miras {\bf (a)} and all variable types {\bf (b)} for S-type stars (green dots), M-type stars (blue squares), and carbon AGB stars (red triangles) samples. Gas expansion velocities plotted against the periods of the Miras {\bf (c)} and all variable types {\bf (d)} for  S-type stars (green dots), M-type stars (blue squares), and carbon AGB stars (red triangles) samples.} 
\label{plot_per}
\end{figure*}

Another indication of this can be seen in Fig.~\ref{plot_per} where the mass-loss rate (a and b) and expansion velocity (c and d) are plotted against the stellar variability period. As already discussed by \citet{schoolof01} for the carbon star sample, the mass-loss rate increases with the period of the star for all chemistries. The same trend can be observed for the expansion velocity. If the efficiency for driving a wind by radiation pressure on dust was weaker in the S-type stars, due to, e.g., a low dust-to-gas ratio or a dust type incapable of driving a wind, the S-type stars would occupy a different parameter space than the M-type stars and the carbon stars in these plots. However, it is not possible to distinguish between the three different types.

\subsection{Dust-to-gas ratio}
The average of the derived dust-to-gas mass-loss-rate ratios is about 2.8\,$\times$\,10$^{-3}$ for the S-type AGB stars in Table~\ref{dust_results1} [V386 Cep excluded; this is likely the same type as GX Mon discussed in \citet{ramsetal08}]. \citet{groeetal99} derived dust-to-gas ratios of 48 M-type stars and found an average of 5.8\,$\times$\,10$^{-3}$, and \citet{ramsetal08} derived dust-to-gas ratios of four high-mass-loss-rate M-type stars resulting in, on average, 2.8\,$\times$\,10$^{-3}$. For the carbon stars the average dust-to-gas ratios might be slightly lower [2.5\,$\times$\,10$^{-3}$, \citet{groeetal98}, 2\,$\times$\,10$^{-3}$, \citet{ramsetal08}]. We conclude that the apparently low dust content in S-type AGB stars reflects the fact that they, for the most part, are low mass-loss rate objects. Their dust-to-gas ratios seem to be in agreement with what is derived for AGB stars of other chemical types, implying that the dust formation efficiency is similar in all three chemical types.


\subsection{Circumstellar SiO abundances and constraints on chemical models}
\label{ss:dis:sio}

The SiO abundances derived for the sample of S-type AGB stars range between $4\times 10^{-7}$ and $1.4\times 10^{-4}$. All stars (except RZ Sgr, which has the highest wind density) have abundances above or well above the expected thermal equilibrium value for C/O\,=\,1 ($6.4\times 10^{-7}$, \citet{cher06}). The median value of $6\times 10^{-6}$ is almost an order of magnitude larger than the equilibrium value. \citet{cher06} derives an SiO abundance at 5 $r_{\star}$ of $3.3\times 10^{-5}$ for C/O\,=\,1 in a non-equilibrium shock-chemistry model. The SiO abundance derived in a shock-chemistry model is sensitive to the specific parameters of the star, such as the pulsational period and the shock velocity. The periods of the S-type AGB stars in our sample range between about 100\,$-$\,600 days but we find no obvious correlation with the derived SiO abundance. However, in a real, dynamic stellar atmosphere the chemistry is most likely much more complicated than in the present chemical models, and simple dependencies on single parameters, like the period, may not exist. We find no correlation between the LRS class and the derived SiO abundance for the S-type stars.

The derived circumstellar SiO abundances for the sample of S-type AGB stars are compared to those obtained for carbon stars \citep{schoetal06} and M-type AGB stars \citep{delgetal03} in Fig.~\ref{plot_sio2}. The three distributions are very similar suggesting that the abundance of SiO in a circumstellar chemistry is not very sensitive to the C/O ratio. In Fig.~\ref{plot_sio} the estimated abundances are plotted against a measure of the density of the wind ($\dot{M}/v_{\rm{e}}$). For a specific density the SiO abundance can vary by up to two orders of magnitude. From the $\chi^{2}$-analysis we find that the abundances derived when also the size of the emitting region is a free parameter, are determined within a factor of $\approx$\,4. Eq.~\ref{re} is derived using a relatively large number of stars with a large range in mass-loss rates and might give a statistically more accurate abundance. Similar investigations of abundances of circumstellar molecules, for which the results are more in line with what would be expected in equilibrium chemistry (e.g., SiS and HCN), show a smaller range in the derived abundances for different stars \citep{schoetal07,schoolof08}.

For AGB stars with low density envelopes, such as the majority of the stars discussed here, condensation of SiO molecules onto dust grains is not very effective and most probably not the cause for the observed scatter in the derived SiO abundances (see Sect.~\ref{ss:dis:dust}). At chemical equilibrium a spread in the abundances is also to be expected depending on the C/O ratio and the temperature of the star. Around C/O\,=\,1, the change in the amount of SiO formed is rather drastic, and between 0.95 and 1.05 the amount of SiO can change with up to two orders of magnitude \citep{mark00}. If the temperature is varied between 2200 and 2600\,K for a given C/O-ratio, the SiO abundance will change less than an order of magnitude\footnote{astrochemistry.net-SiO, H2, 19 Jan 2009}. For the S-type stars alone it is not possible to conclude whether the spread in the derived abundances is indicative of a non-equilibrium chemistry or due to a spread in the C/O-ratio around 1. However, given the results for all three chemical types, we conclude that the spread in the derived circumstellar SiO abundances most probably is real and indicative of a shock-chemistry in the formation zone of SiO. 

The derived SiO envelope sizes (see Table~\ref{abund_best} and Fig.~\ref{size}) can be used to test photochemical models of the circumstellar envelope. Using a simple photochemical model \citep[][and references therein]{lindetal00,delgetal03} with typical stellar, circumstellar and dust parameters for our sample of S-type AGB stars, and adopting an unshielded photodissociation rate of 2.5\,$\times$\,10$^{-10}$\,s$^{-1}$ \citep{delgetal03}, we obtain the relation between SiO envelope size ($r_{\mathrm e}$) and density measure ($\dot{M}/v_{\mathrm{e}}$) shown in Fig.~\ref{size} (dashed line). These predictions agree well with the observed values and the relation described by Eq.~\ref{re} (solid line in Fig.~\ref{size}) used in the abundance estimates.


\subsection{SiO maser emission}
There are several indications that SiO maser emission originates from close to the stellar photosphere \citep{reidmora81}. The energy of the masing state is about 1800\,K ($v$=1) or higher, and VLBA observations confirm that the masers are formed within a few stellar radii of the star \citep[see for instance][and references therein]{cottetal06}. This corresponds to the region where the dust is formed \citep{salp74,dancetal94,reidment97}. Observations of SiO masers can therefore provide information on whether there is SiO present in the dust formation region, and hence whether it can be one of the constituents of the dust formed. The non-detection of SiO maser emission toward carbon stars is thought to be indicating that the SiO molecules are formed further out in the wind than in M-type AGB stars.

SiO maser emission has been searched for in 30 of our sample stars, and has previously been detected in nine \citep{joriknap98}. The results of our SiO($J$\,=\,2$\rightarrow$\,1, $v$=1) observations are given in Table~\ref{SiOintensitiesm1}, and the detected lines are shown in Fig.~\ref{rand_siom}. In four stars the SiO maser has previously not been detected, and in another four stars we can not confirm previous detections (R Cyg, R Lyn, Y Lyn and EP Vul), most likely a result of time variability. In Table~\ref{SiOintensitiesm1} we give distance independent upper limits for the sources observed. Unfortunately the reason for the non-detections is not possible to identify but time variability may play a role.

Out of the 12 S-type stars in our sample classified as showing silicate emission in their LRS spectra, eight were observed and six were detected. Nine stars are classified as being featureless and out of these, six were observed and one was detected. Out of the three stars classified as not showing any IR excess, two were observed and none was detected, and finally, out of the 13 unclassified stars, seven were observed and two were detected in the SiO($J$\,=\,2$\rightarrow$\,1, $v$=1) maser line. These results are in line with a situation where the silicate stars have a C/O-ratio slightly less than one, and hence have enough SiO in their photospheres to produce SiO masers, while the others, in general, have C/O-ratios slightly larger than one, and hence too low photospheric SiO abundances to produce SiO masers.



\subsection{SiO chemistry and implications for dust formation}
\label{ss:dis:dust}

The formation of SiO in AGB winds with shocks is discussed by \citet{cher06} for different C/O-ratios. It is described how the formation of SiO is linked to the presence of OH and how the lack of OH in the inner wind of carbon stars explain their inability to form silicate dust despite their high SiO abundances further out in the CSE. 

\begin{figure}[h]
\raggedright 
{\includegraphics[width=\columnwidth]{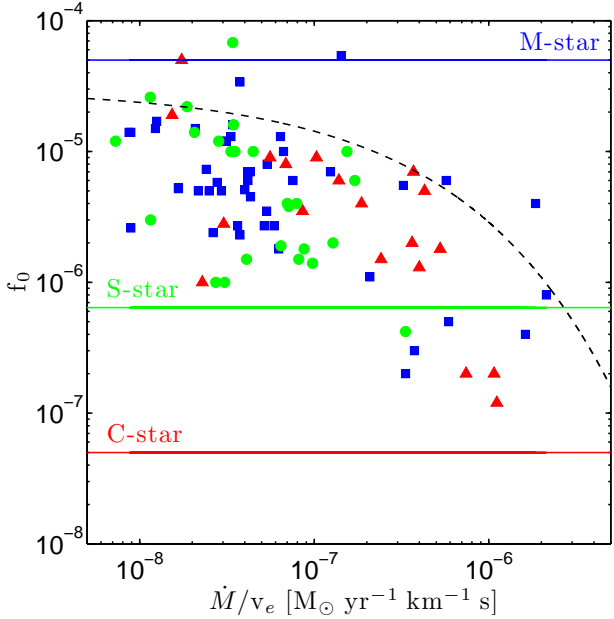}}
\caption{Circumstellar SiO fractional abundances ($f_{0}$) as a function of wind density ($\dot{M}/v_{\rm{e}}$) for S-type (green dots), M-type (blue squares) \citep{delgetal03}, and carbon stars (red triangles) \citep{schoetal06}. The horizontal lines mark the abundances predicted from equilibrium chemistries \citep{cher06}. The dashed line shows the expected post-condensation abundance, $f(\infty)$, (scaled to 3.0$\times$10$^{-5}$, roughly the expected fractional abundance at 5 $r_{\star}$ for low mass-loss rates when C/O=1) from a model including adsorption of SiO onto dust grains \citep[see][for details]{delgetal03}.}
\label{plot_sio}
\end{figure}

For the S-type AGB stars, the situation is more complicated. The exact C/O-ratio is not known for the stars in our sample, but it is most likely a spread around 1. The C/O-ratio of course has implications for both the chemistry and the dust formed. \citet{cher06} finds that the chemistry of the inner winds of stars with C/O\,=\,0.98 is very similar to the M-type chemistry, while the chemistry of stars with C/O\,=\,1.01 is very similar to that of carbon stars. In our comparison (Figs~\ref{plot_sio2} and \ref{plot_sio}) it is not possible to distinguish between the different chemical types in terms of their circumstellar SiO abundances. Given also the similarity of the mass-loss-rate distributions for the different chemical types, the most likely explanation is that the S-type stars in our sample have a spread in their C/O-ratio where some stars are more M-type like (forming silicate dust) and some are more carbon-star like (forming carbon dust). However, we are not able to exclude a scenario where two components of dust have formed in stars with a C/O-ratio close to unity.

For M-type AGB stars \citep{delgetal03} and carbon stars \citep{schoetal06} there is a clear trend that the circumstellar SiO abundance gets lower as the density of the wind increases (Fig.~\ref{plot_sio}), indicative of adsorption of SiO onto dust grains. For the sample of S-type AGB stars only an indication of such a trend may be seen, possibly due to the lack of high mass-loss rate objects. The dashed line in Fig.~\ref{plot_sio} shows a depletion curve based on a simple model presented in \citet{delgetal03}. A relatively large scatter around the curve is expected since the condensation is sensitive to stellar, circumstellar, and dust characteristics. Further support for a depletion scenario comes from recent interferometric observations.
\citet{schoetal06a} modelled interferometric observations of SiO($J$\,=\,5$\rightarrow$\,4) emission and infrared observations of rovibrational transitions of the extreme carbon star IRC+10216, and found that a two-component radial abundance distribution (a compact high abundance, pre-condensation, component combined with a more extended low abundance, post-condensation, component) is needed in order to explain the observations. Models of interferometric observations of M-type AGB stars \citep{schoetal04} also show that a two-component radial abundance distribution gives a better fit to the observed data. The observations at hand for the S-type AGB stars do not allow a proper investigation of a scenario in which SiO molecules are incorporated into grains (see discussion in \citet{schoetal07}). Instead, the abundances derived here should be considered as post-condensation abundances.


\section{Conclusions}
\label{s:conc}

We have modelled multi-transitional CO radio line data from 40 S-type AGB stars. We have also modelled multi-transitional SiO data of 26 stars from the same sample. The results are compared to similar samples of M-type AGB stars \citep{olofetal02} and carbon stars \citep{schoolof01}. We arrive at the following conclusions:

\begin{itemize}

\item{We find that the mass-loss rate distributions are very similar for the three chemical types (Fig.~\ref{plot_co}a), and so are the relations between mass-loss rate and expansion velocity of the stellar wind (Fig.~\ref{plot_co}c). Further, it is not possible to distinguish between the different chemical types when examining how the mass-loss rate correlates with the pulsation period of the star (Fig.~\ref{plot_per}). The most likely explanation for the observed trends is a mass loss that is driven by the same mechanism(s) in all three chemical types. The only apparent difference is that the carbon star CSEs have higher gas expansion velocities (on average), most likely an effect of a higher acceleration efficiency of the carbon grains.}

\item{The derived dust-to-gas mass-loss-rate ratios for 11 of the sample stars show an agreement with what has previously been found for M-type AGB stars and carbon stars, implying that the dust formation efficiency is similar in all three chemical types.}

\item{The median value of the estimated circumstellar SiO fractional abundances in S-type AGB stars is almost more than an order of magnitude higher than predicted by thermal equilibrium chemistry, suggesting that shock-induced non-equilibrium chemical processes are important in regulating the chemistry in the inner wind.}

\item{The derived circumstellar SiO abundances for the S-type AGB stars range from 4\,$\times$\,10$^{-7}$ to 1.4\,$\times$\,10$^{-4}$. For a specific wind density ($\dot{M}/v_{\mathrm{e}}$) there is a scatter of almost two orders of magnitudes, in accordance with what has previously been found for the other chemical types. In terms of their distribution of SiO circumstellar abundances, it is not possible to distinguish between the three different chemical types, and we propose that, although there is a large scatter, the circumstellar SiO abundance is independent of the C/O-ratio for a given mass-loss rate and expansion velocity. 
The formation efficiency of SiO in a shock chemistry will be sensitive to other specific parameters of the star, like pulsational period and shock velocity. We believe that the scatter obtained in the SiO abundance estimates is real and indicative of a shock chemistry in the outer atmosphere.}

\item{Previous analysis of M-type AGB stars \citep{delgetal03} and carbon stars \citep{schoetal06} show a clear trend in that the circumstellar SiO abundance decreases as the density of the stellar wind increases. The trend is indicative of adsorption of SiO onto dust grains. The same trend can be suspected for the sample of S-type AGB stars, however, the number of high-mass-loss-rate S-type AGB stars is low and no firm conclusion can be drawn.}

\end{itemize}

Despite the limitations due to the selection criteria of the samples compared, we believe that our conclusions apply to AGB stars in general.

\begin{acknowledgements}
The authors acknowledge support from the Swedish Research Council. This work has benefited from research funding from the European Community's Sixth Framework Programme under RadioNet contract R113CT 2003 5158187 and R113CT 2003 5058187.
This article made use of data obtained through the JCMT archive as Guest User 
at the Canadian Astronomy Data Center, which is operated by the Dominion
Astrophysical Observatory for the National Research Council of Canada's
Herzberg Institute of Astrophysics. 
\end{acknowledgements}

\bibliographystyle{aa}
\bibliography{paper}

\clearpage

\appendix

\section{New CO spectra}
\label{a:cospectra}

\begin{table}
\caption{Observational results of circumstellar CO radio line emission.}
\label{COintensities1}
$$
\begin{array}{p{0.15\linewidth}ccccccc}
\hline
\noalign{\smallskip}
\multicolumn{1}{l}{{\mathrm{Source}}} &
\multicolumn{1}{c}{{\mathrm{Transition}}} &&
\multicolumn{1}{c}{{I_{\rm{mb}}}} &&
\multicolumn{1}{c}{{T_{\rm{mb}}}} &&
 \multicolumn{1}{c}{\rm{Telescope}} \\ 
& &&
\multicolumn{1}{c}{\mathrm{[K \ km \ s^{-1}]}} &&
\multicolumn{1}{c}{[\mathrm{K}]} &&
\multicolumn{1}{c}{\rm{or \ Reference}}
\\
\noalign{\smallskip}
\hline
\noalign{\smallskip}
\object{R And} & \rm{CO(1-0)} && \phantom{11}5.7 && 0.44 && \rm{BL94} \\
 & \rm{CO(1-0)} && \phantom{1}34.9 && 2.4\phantom{1} && \rm{I\phantom{^{a}}} \\
 & \rm{CO(2-1)} && 135.5 && 9.6\phantom{1} && \rm{I\phantom{^{a}}} \\
 & \rm{CO(2-1)} && \phantom{1}32.0 && 2.5\phantom{1} && \rm{J^{a}} \\
 & \rm{CO(3-2)} && \phantom{1}43.0 && 3.3\phantom{1} && \rm{J^{a}} \\
 \\
\object{W And} & \rm{CO(1-0)} && \phantom{11}1.4 && 0.12 && \rm{BL94} \\
 & \rm{CO(1-0)} && \phantom{11}3.8 && 0.4\phantom{1} && \rm{O\phantom{^{a}}} \\
 & \rm{CO(1-0)} && \phantom{11}9.5 && 0.9\phantom{1} && \rm{I\phantom{^{a}}} \\
 & \rm{CO(2-1)} && \phantom{1}55.3 && 4.7\phantom{1} && \rm{I\phantom{^{a}}} \\
 & \rm{CO(3-2)} && \phantom{11}7.0 && 0.6\phantom{1} && \rm{Y95} \\
 \\
\object{Z Ant} & \rm{CO(2-1)} && \phantom{11}0.7 && 0.07 && \rm{JK98} \\
 & \rm{CO(3-2)} && \phantom{11}2.0 && 0.6\phantom{1} && \rm{A\phantom{^{a}}} \\
 \\
\object{VX Aql} &  \rm{CO(1-0)} && \phantom{11}0.7 && 0.05 && \rm{SL95} \\
 \\
\object{W Aql} & \rm{CO(1-0)} &&  \phantom{1}28.4 && 1.3\phantom{1} && \rm{N92} \\
 & \rm{CO(2-1)} && \phantom{1}54.8 && 2.1\phantom{1} && \rm{K98} \\
 & \rm{CO(3-2)} && \phantom{1}93.2 && 3.5\phantom{1} && \rm{K98} \\
 & \rm{CO(3-2)} && 136.2 && 5.0\phantom{1} && \rm{A\phantom{^{a}}} \\
 & \rm{CO(4-3)} && 199.5 && 7.4\phantom{1} && \rm{J^{a}} \\	
 \\
\object{AA Cam} & \rm{CO(2-1)} && \phantom{11}1.8 && 0.06 && \rm{SL95} \\
 & \rm{CO(3-2)} && \phantom{11}0.8 && 0.17 && \rm{J\phantom{^{a}}} \\
 \\
\object{T Cam} & \rm{CO(1-0)} && \phantom{11}2.0 && 0.3\phantom{1} &&  \rm{I\phantom{^{a}}} \\
 & \rm{CO(2-1)} && \phantom{11}8.1 && 1.2\phantom{1} && \rm{I\phantom{^{a}}} \\
  \\
\object{S Cas} & \rm{CO(1-0)} && \phantom{1}13.8 && 0.4\phantom{1} && \rm{O\phantom{^{a}}} \\
 & \rm{CO(1-0)} && \phantom{1}38.2 && 1.1\phantom{1} && \rm{I\phantom{^{a}}} \\
 & \rm{CO(2-1)} && 157.3 && 4.8\phantom{1} && \rm{I\phantom{^{a}}} \\
 & \rm{CO(3-2)} && \phantom{1}31.0 && 1.1\phantom{1} && \rm{J^{a}} \\
 \\	
\object{V365 Cas}	& \rm{CO(2-1)} && \phantom{11}0.7 && 0.07 && \rm{SL95} \\
 \\
\object{WY Cas}  & \rm{CO(1-0)} && \phantom{11}3.5 && 0.12 && \rm{O\phantom{^{a}}} \\
 & \rm{CO(1-0)} && \phantom{1}12.8 && 0.6\phantom{1} && \rm{I\phantom{^{a}}} \\
 & \rm{CO(2-1)} && \phantom{1}64.4 && 2.9\phantom{1} && \rm{I\phantom{^{a}}} \\
 \\
\object{AM Cen}  & \rm{CO(1-0)} && \phantom{11}0.6 && 0.08 && \rm{SL95} \\
 & \rm{CO(3-2)} && \phantom{11}1.1 && 0.24 && \rm{A\phantom{^{a}}} \\
 \\
\object{TT Cen}  & \rm{CO(1-0)} && \phantom{11}1.9 && 0.05 && \rm{SL95} \\
 & \rm{CO(3-2)} && \phantom{1}13.9 && 0.5\phantom{1} && \rm{A\phantom{^{a}}} \\
 \\
\object{UY Cen} & \rm{CO(2-1)} && \phantom{11}1.0 && 0.07 && \rm{SL95} \\
 & \rm{CO(3-2)} && \phantom{11}1.6 && 0.1\phantom{1} && \rm{A\phantom{^{a}}} \\
 \\
\object{V386 Cep} & \rm{CO(2-1)} && \phantom{11}8.4 && 0.3\phantom{1} && \rm{I\phantom{^{a}}} \\
\\
\object{T Cet} & \rm{CO(2-1)} && \phantom{11}2.0 && 0.28 && \rm{S\phantom{^{a}}} \\
 & \rm{CO(2-1)} && \phantom{11}3.7 && 0.4\phantom{1} && \rm{J^{a}} \\
 & \rm{CO(3-2)} && \phantom{11}6.1 && 0.7\phantom{1} && \rm{A\phantom{^{a}}} \\
 & \rm{CO(3-2)} && \phantom{11}7.2 && 0.8\phantom{1} && \rm{J^{a}} \\
 & \rm{CO(4-3)} && \phantom{11}9.5 && 0.97 && \rm{J^{a}} \\
 & \rm{CO(6-5)} && \phantom{1}10.0 && 1.1\phantom{1} && \rm{J^{a}} \\

\noalign{\smallskip}
\hline
\end{array}
$$
\end{table}
\begin{table}
\caption{Observational results of circumstellar CO radio line emission (continued).}
\label{COintensities2}
$$
\begin{array}{p{0.15\linewidth}ccccccc}
\hline
\noalign{\smallskip}
\multicolumn{1}{l}{{\mathrm{Source}}} &
\multicolumn{1}{c}{{\mathrm{Transition}}} &&
\multicolumn{1}{c}{{I_{\rm{mb}}}} &&
\multicolumn{1}{c}{{T_{\rm{mb}}}} &&
 \multicolumn{1}{c}{\rm{Telescope}} \\ 
& &&
\multicolumn{1}{c}{\mathrm{[K \ km \ s^{-1}]}} &&
\multicolumn{1}{c}{[\mathrm{K}]} &&
\multicolumn{1}{c}{\rm{or \ reference}}
\\
\noalign{\smallskip}
\hline
\noalign{\smallskip}
\object{AA Cyg} & \rm{CO(1-0)} && \phantom{11}1.5 && \phantom{1}0.11 && \rm{BL94} \\
 & \rm{CO(1-0)} && \phantom{11}5.7 && \phantom{1}0.7\phantom{1} && \rm{I\phantom{^{a}}} \\
 & \rm{CO(2-1)} && \phantom{1}17.4 && \phantom{1}2.3\phantom{1} && \rm{I\phantom{^{a}}} \\
  \\
\object{AD Cyg} & \rm{CO(1-0)} && \phantom{11}0.9 && \phantom{1}0.08 && \rm{I\phantom{^{a}}} \\
 & \rm{CO(2-1)} && \phantom{11}3.8 && \phantom{1}0.31 && \rm{I\phantom{^{a}}} \\
 \\
\object{R Cyg} & \rm{CO(1-0)} && \phantom{11}2.5 && \phantom{1}0.15 && \rm{BL94} \\	
 & \rm{CO(1-0)} && \phantom{11}4.4 && \phantom{1}0.3\phantom{1} && \rm{O\phantom{^{a}}} \\
 & \rm{CO(1-0)} && \phantom{1}16.4 && \phantom{1}1.0\phantom{1} && \rm{I\phantom{^{a}}} \\
 & \rm{CO(2-1)} && \phantom{1}70.2 && \phantom{1}4.2\phantom{1} && \rm{I\phantom{^{a}}} \\
 & \rm{CO(3-2)} && \phantom{1}14.6 && \phantom{1}2.3\phantom{1} && \rm{S95} \\
 \\
\object{$\chi$ Cyg}	& \rm{CO(1-0)} && \phantom{1}27.2 && \phantom{1}1.8\phantom{1} && \rm{O\phantom{^{a}}} \\
 & \rm{CO(1-0)} && \phantom{1}70.9 && \phantom{1}6.0\phantom{1} && \rm{I\phantom{^{a}}} \\
 & \rm{CO(2-1)} && \phantom{1}60.2 && \phantom{1}4.2\phantom{1} && \rm{J^{a}} \\
 & \rm{CO(2-1)} && 265.8 && 19.8\phantom{1} && \rm{I\phantom{^{a}}} \\
 & \rm{CO(3-2)} && 119.5 && \phantom{1}8.3\phantom{1} && \rm{J\phantom{^{a}}} \\
 & \rm{CO(4-3)} && 134.6 && \phantom{1}9.6\phantom{1} && \rm{J^{a}} \\
 \\
\object{TV Dra}	& \rm{CO(1-0)} && \phantom{11}0.6 && \phantom{1}0.02 && \rm{O\phantom{^{a}}} \\
 & \rm{CO(1-0)} && \phantom{11}1.5 && \phantom{1}0.16 && \rm{I\phantom{^{a}}} \\
 & \rm{CO(2-1)} && \phantom{11}4.1 && \phantom{1}0.6\phantom{1} && \rm{I\phantom{^{a}}} \\
 \\
\object{DY Gem} & \rm{CO(1-0)} && \phantom{11}4.4 && \phantom{1}0.3\phantom{1} && \rm{I\phantom{^{a}}} \\
 & \rm{CO(2-1)} && \phantom{1}17.1 && \phantom{1}1.3\phantom{1} && \rm{I\phantom{^{a}}} \\
 & \rm{CO(3-2)} && \phantom{11}2.6 && \phantom{1}0.35 && \rm{A\phantom{^{a}}} \\
 \\
\object{R Gem} & \rm{CO(1-0)} && \phantom{11}2.4 && \phantom{1}0.3\phantom{1} && \rm{O\phantom{^{a}}} \\
 & \rm{CO(1-0)} && \phantom{11}9.7 && \phantom{1}1.3\phantom{1} && \rm{I\phantom{^{a}}} \\
 & \rm{CO(2-1)} && \phantom{1}34.1 && \phantom{1}5.2\phantom{1} && \rm{I\phantom{^{a}}} \\
 & \rm{CO(3-2)} && \phantom{11}6.5 && \phantom{1}0.97 && \rm{A\phantom{^{a}}} \\
 \\
\object{ST Her} & \rm{CO(1-0)} && \phantom{11}1.5 && \phantom{1}0.2\phantom{1} && \rm{O\phantom{^{a}}} \\
 & \rm{CO(1-0)} && \phantom{11}4.9 && \phantom{1}0.24 && \rm{KJ94} \\
 & \rm{CO(2-1)} && \phantom{1}17.4 && \phantom{1}1.05 && \rm{KJ94} \\
 & \rm{CO(3-2)} && \phantom{11}5.1 && \phantom{1}0.34 && \rm{K98} \\
 \\
\object{RX Lac} &  \rm{CO(1-0)} && \phantom{11}2.0 && \phantom{1}0.3\phantom{1} && \rm{GJ98} \\
 \\
\object{GI Lup} & \rm{CO(3-2)} && \phantom{11}4.6 && \phantom{1}0.3\phantom{1} && \rm{A\phantom{^{a}}} \\
 \\	
\object{R Lyn} & \rm{CO(1-0)} && \phantom{11}0.3 && \phantom{1}0.03 && \rm{BL94} \\
 & \rm{CO(1-0)} && \phantom{11}0.8 && \phantom{1}0.1\phantom{1} && \rm{O\phantom{^{a}}} \\
 & \rm{CO(1-0)} && \phantom{11}2.0 && \phantom{1}0.18 && \rm{I\phantom{^{a}}} \\
 & \rm{CO(2-1)} && \phantom{11}5.8 && \phantom{1}0.5\phantom{1} && \rm{I\phantom{^{a}}} \\
 & \rm{CO(2-1)} && \phantom{11}0.9 && \phantom{1}0.08 && \rm{JK98} \\
 \\
\object{Y Lyn} & \rm{CO(1-0)} && \phantom{11}4.1 && \phantom{1}0.3\phantom{1} && \rm{O\phantom{^{a}}} \\
 & \rm{CO(1-0)} && \phantom{1}14.3 && \phantom{1}1.1\phantom{1} && \rm{I\phantom{^{a}}} \\
 & \rm{CO(2-1)} && \phantom{1}56.4 && \phantom{1}4.4\phantom{1} && \rm{I\phantom{^{a}}} \\
 \\
\object{S Lyr} & \rm{CO(1-0)} && \phantom{11}1.9 && \phantom{1}0.1\phantom{1} && \rm{O\phantom{^{a}}} \\
 & \rm{CO(3-2)} && \phantom{11}6.9 && \phantom{1}0.32 && \rm{A\phantom{^{a}}} \\
\smallskip \\ 	
\object{FU Mon} & \rm{CO(1-0)} && \phantom{11}3.3 && \phantom{1}0.7\phantom{1} && \rm{I\phantom{^{a}}} \\
 & \rm{CO(2-1)} && \phantom{11}7.0 && \phantom{1}2.0\phantom{1} && \rm{I\phantom{^{a}}} \\
 & \rm{CO(3-2)} && \phantom{11}1.3 && \phantom{1}0.3\phantom{1} && \rm{A\phantom{^{a}}} \\
 \\
\object{RZ Peg} & \rm{CO(2-1)} && \phantom{11}3.5 && \phantom{1}0.21 && \rm{SL95} \\
\noalign{\smallskip}
\hline
\end{array}
$$
\end{table}
\begin{table}
\caption{Observational results of circumstellar CO radio line emission (continued).}
\label{COintensities3}
$$
\begin{array}{p{0.2\linewidth}ccccccc}
\hline
\noalign{\smallskip}
\multicolumn{1}{l}{{\mathrm{Source}}} &
\multicolumn{1}{c}{{\mathrm{Transition}}} &&
\multicolumn{1}{c}{{I_{\rm{mb}}}} &&
\multicolumn{1}{c}{{T_{\rm{mb}}}} &&
 \multicolumn{1}{c}{\rm{Telescope}} \\ 
& &&
\multicolumn{1}{c}{\mathrm{[K \ km \ s^{-1}]}} &&
\multicolumn{1}{c}{[\mathrm{K}]} &&
\multicolumn{1}{c}{\rm{or \ reference}}
\\
\noalign{\smallskip}
\hline
\noalign{\smallskip}
\object{RT Sco} & \rm{CO(1-0)} && \phantom{1}3.8\phantom{:} && 0.20\phantom{:} && \rm{SL95} \\
 & \rm{CO(3-2)} && 32.6\phantom{:} && 1.8\phantom{1:} && \rm{A} \\
 \\
\object{ST Sco} & \rm{CO(1-0)} && \phantom{1}0.6\phantom{:} && 0.06\phantom{:} && \rm{BL94} \\
 & \rm{CO(1-0)} && \phantom{1}1.4\phantom{:} && 0.14\phantom{:} && \rm{SL95} \\
 & \rm{CO(3-2)} && \phantom{1}4.5\phantom{:} && 0.5\phantom{1:} && \rm{A} \\
 \\
\object{RZ Sgr} & \rm{CO(1-0)} && \phantom{1}6.4\phantom{:} && 0.38\phantom{:} && \rm{S} \\
 & \rm{CO(2-1)} && 10.7\phantom{:} && 0.87\phantom{:} && \rm{S} \\
 & \rm{CO(3-2)} && 11.7\phantom{:} && 1.0\phantom{1:} && \rm{A} \\
 \\
\object{ST Sgr} &  \rm{CO(1-0)} && \phantom{1}0.8\phantom{:} && 0.66\phantom{:} && \rm{SL95} \\
 & \rm{CO(3-2)} && \phantom{1}4.9\phantom{:} && 1.0\phantom{1:} && \rm{A} \\
 \\
\object{T Sgr} & \rm{CO(1-0)} && \phantom{1}1.5\phantom{:} && 0.17\phantom{:} && \rm{I} \\
 & \rm{CO(2-1)} && \phantom{1}5.2\phantom{:} && 0.4\phantom{1:} && \rm{I} \\
 & \rm{CO(3-2)} && \phantom{1}2.5\phantom{:} && 0.2\phantom{1:} && \rm{A} \\
 \\ 
\object{EP Vul} & \rm{CO(1-0)} && \phantom{1}0.9\phantom{:} && 0.12\phantom{:} && \rm{BL94} \\
 & \rm{CO(1-0)} && \phantom{1}0.6\phantom{:} && 0.08\phantom{:} && \rm{SL95} \\
 & \rm{CO(3-2)} && \phantom{1}3.1\phantom{:} && 0.35\phantom{:} && \rm{A} \\
 \\
\object{DK Vul} & \rm{CO(1-0)} && \phantom{1}1.3\phantom{:} && 0.04\phantom{:} && \rm{BL94} \\
& \rm{CO(1-0)} && \phantom{1}2.9\phantom{:} && 0.6\phantom{1:} && \rm{I} \\
 & \rm{CO(2-1)} && \phantom{1}9.6\phantom{:} && 1.7\phantom{1:} && \rm{I} \\
 & \rm{CO(3-2)} && \phantom{1}3.6\phantom{:} && 0.7\phantom{1:} && \rm{A} \\
 \\ 
\object{AFGL 2425} & \rm{CO(1-0)} && \phantom{1}3.3\phantom{:} && 0.2\phantom{1:} && \rm{I} \\
 & \rm{CO(2-1)} && 10.5\phantom{:} && 0.75\phantom{:} && \rm{I} \\
 & \rm{CO(3-2)} && \phantom{1}2.8\phantom{:} && 0.25\phantom{:} && \rm{A} \\
 \\
 \object{CSS2 41} & \rm{CO(1-0)} && \phantom{1}2.6: && 0.15: && \rm{I} \\
 & \rm{CO(2-1)} && \phantom{1}4.7: && 0.3\phantom{1}: && \rm{I} \\
 & \rm{CO(3-2)} && \phantom{1}2.5: && 0.1\phantom{1}: && \rm{J} \\	
 \\
\object{IRC--10401}	& \rm{CO(1-0)} && \phantom{1}2.9\phantom{:} && 0.12\phantom{:} && \rm{GJ98} \\
 & \rm{CO(3-2)} && \phantom{1}9.5\phantom{:} && 0.45\phantom{:} && \rm{A} \\
 \\
\noalign{\smallskip}
\hline
\end{array}
$$
A; APEX, I; IRAM, J; JCMT, J$^{a}$; JCMT archival data, O; OSO \\ BL94; \citet{bieglatt94}, Y95; \citet{youn95}, JK98; \citet{joriknap98}, SL95; \citet{sahaliec95}, N92; \citet{nymaetal92}, K98; \citet{knapetal98}, S95; \citet{stanetal95}, KJ94; \citet{kahajura94}, GJ98; \citet{groedejo98}
\end{table}

\begin{figure*}[h]
\raggedright 
{\includegraphics[width=18cm]{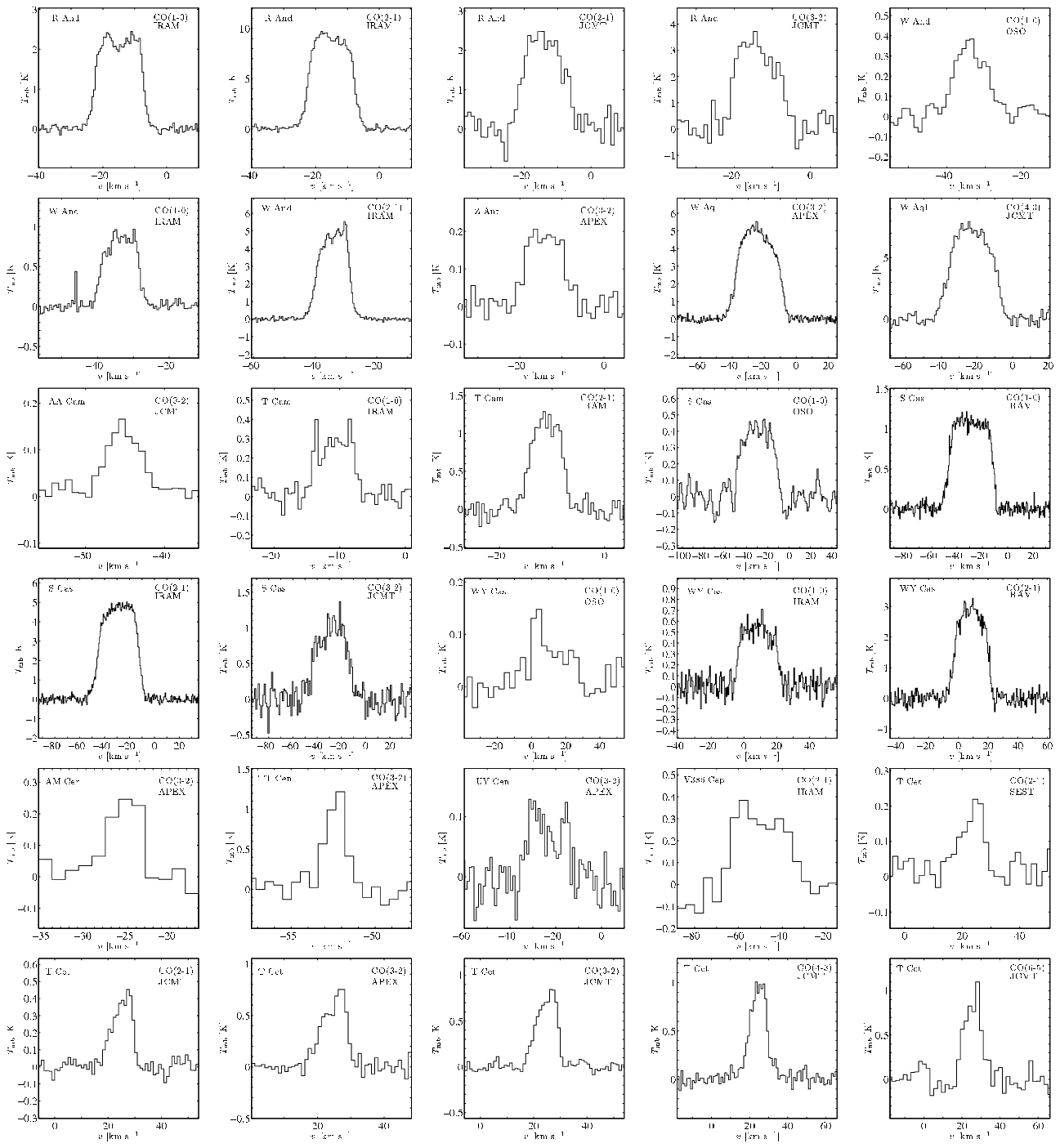}}
\caption{New observations of circumstellar CO line emission. The source name is shown to the upper left, and the transition observed and the telescope used is shown to the upper right of each frame. LSR velocity scales are used.}
\label{rand_co1}
\end{figure*}

\begin{figure*}[h]
\raggedright 
{\includegraphics[width=18cm]{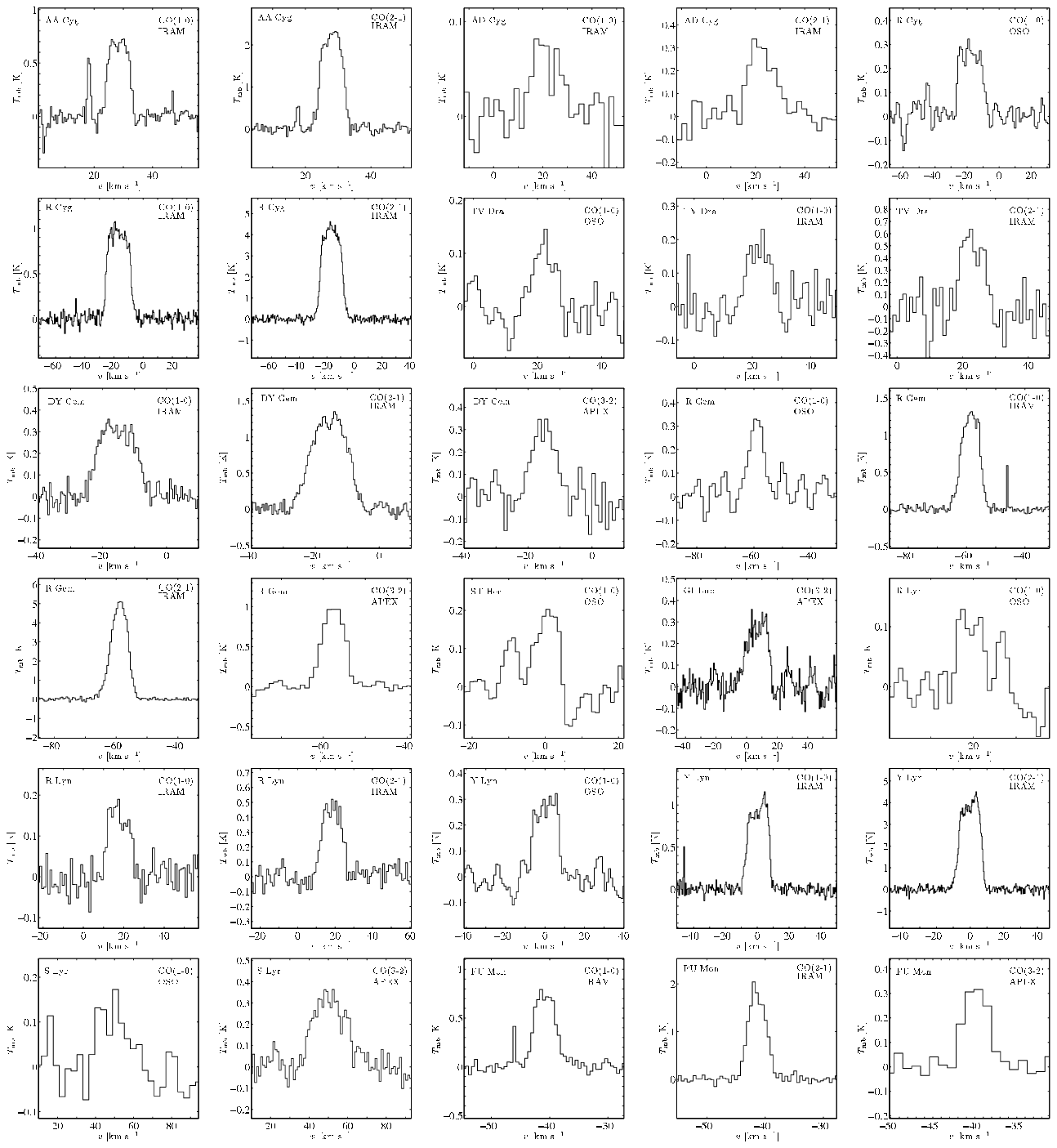}}
\caption{New observations of circumstellar CO line emission. The source name is shown to the upper left, and the transition observed and the telescope used is shown to the upper right of each frame. LSR velocity scales are used.}
\label{rand_co2}
\end{figure*}

\begin{figure*}[h]
\raggedright 
{\includegraphics[width=18cm]{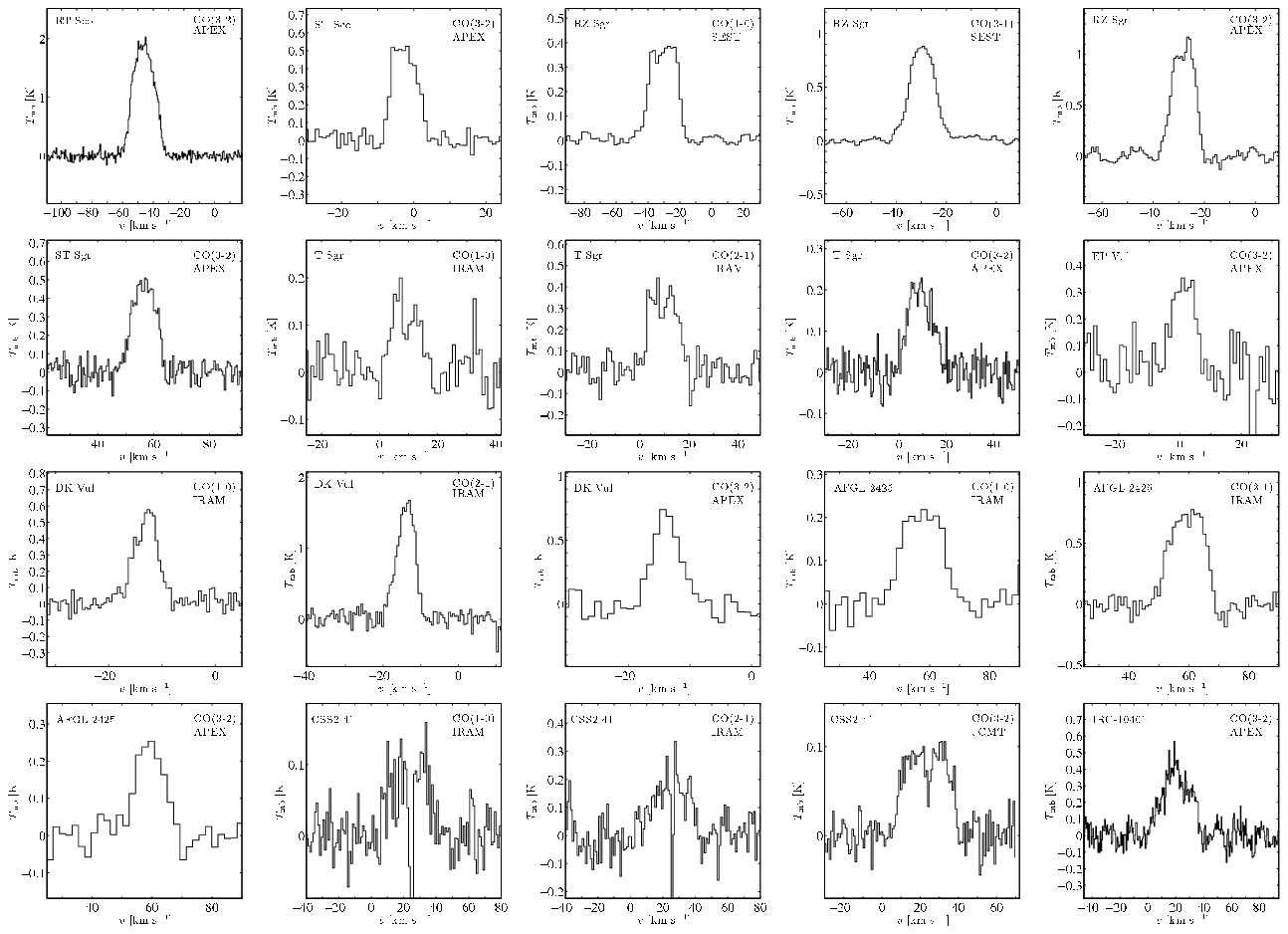}}
\caption{New observations of circumstellar CO line emission. The source name is shown to the upper left, and the transition observed and the telescope used is shown to the upper right of each frame. LSR velocity scales are used.}
\label{rand_co3}
\end{figure*}


\section{New SiO spectra}
\label{a:siospectra}

\begin{table}
\caption{Observational results of circumstellar SiO radio line emission.}
\label{SiOintensities1}
$$
\begin{array}{p{0.15\linewidth}ccccccc}
\hline
\noalign{\smallskip}
\multicolumn{1}{l}{{\mathrm{Source}}} &
\multicolumn{1}{c}{{\mathrm{Transition}}} &&
\multicolumn{1}{c}{{I_{\rm{mb}}}} &&
\multicolumn{1}{c}{{T_{\rm{mb}}}} &&
 \multicolumn{1}{c}{\rm{Telescope}} \\ 
& &&
\multicolumn{1}{c}{\mathrm{[K \ km \ s^{-1}]}} &&
\multicolumn{1}{c}{[\mathrm{K}]} &&
\multicolumn{1}{c}{\rm{or \ Reference}}
\\
\noalign{\smallskip}
\hline
\noalign{\smallskip}
\object{R And} & \rm{SiO(2-1)} && \phantom{1}0.9 && 0.06 && \rm{O} \\	
 & \rm{SiO(2-1)} && \phantom{1}2.5 && 0.22 && \rm{I} \\
 & \rm{SiO(5-4)} && 13.4 && 1.12 && \rm{I} \\
 & \rm{SiO(6-5)} && \phantom{1}3.2 && 0.31 && \rm{J} \\
 & \rm{SiO(8-7)} && \phantom{1}3.5 && 0.39 && \rm{A} \\
 \\
\object{W And} & \rm{SiO(2-1)} && \phantom{1}1.7 && 0.25 && \rm{I} \\
 & \rm{SiO(5-4)} && \phantom{1}6.8 && 1.1\phantom{1} && \rm{I} \\
 \\
\object{W Aql} & \rm{SiO(2-1)} && \phantom{1}2.6 && 0.08 && \rm{O} \\
 & \rm{SiO(3-2)} && \phantom{1}3.9 && 0.15 && \rm{S/B98} \\
 & \rm{SiO(8-7)} && 10.1 && 0.52 && \rm{A} \\
 & \rm{SiO(8-7)} && 16.9 && 0.84 && \rm{J} \\
 \\
\object{S Cas} & \rm{SiO(2-1)} && \phantom{1}1.7 && 0.06 && \rm{O} \\
 & \rm{SiO(2-1)} && \phantom{1}6.1 && 0.22 && \rm{I} \\
 & \rm{SiO(5-4)} && 35.8 && 1.5\phantom{1} && \rm{I} \\
 \\	
\object{WY Cas}  & \rm{SiO(2-1)} && \phantom{1}0.6 && 0.04 && \rm{I} \\
 & \rm{SiO(5-4)} && \phantom{1}3.3 && 0.2\phantom{1} && \rm{I} \\
 & \rm{SiO(8-7)} && \phantom{1}0.9 && 0.11 && \rm{J} \\
 \\
\object{V386 Cep} & \rm{SiO(2-1)} && \phantom{1}1.2 && 0.07 && \rm{I} \\
 & \rm{SiO(5-4)} && \phantom{1}4.3 && 0.26 && \rm{I} \\
 & \rm{SiO(8-7)} && \phantom{1}2.1 && 0.13 && \rm{J} \\
\\
\object{T Cet} & \rm{SiO(8-7)} && \phantom{1}0.7 && 0.03 && \rm{A} \\
\\
\object{AA Cyg} & \rm{SiO(2-1)} && \phantom{1}0.3 && 0.05 && \rm{I} \\
 & \rm{SiO(5-4)} && \phantom{1}2.0 && 0.3\phantom{1} && \rm{I} \\
  \\
\object{R Cyg} & \rm{SiO(2-1)} && \phantom{1}0.3 && 0.03 && \rm{O} \\
 & \rm{SiO(2-1)} && \phantom{1}1.3 && 0.14 && \rm{I} \\
 & \rm{SiO(5-4)} && \phantom{1}6.3 && 0.65 && \rm{I} \\
 & \rm{SiO(8-7)} && \phantom{1}2.0 && 0.20 && \rm{J} \\
 \\
\object{$\chi$ Cyg}	& \rm{SiO(2-1)} && \phantom{1}4.5 && 0.47 && \rm{O} \\
 & \rm{SiO(2-1)} && 23.1 && 1.94 && \rm{I} \\
 & \rm{SiO(5-4)} && 72.3 && 6.7\phantom{1} && \rm{I} \\
 & \rm{SiO(8-7)} && 38.1 && 3.56 && \rm{J} \\
 \\
\object{TV Dra}	& \rm{SiO(2-1)} && \phantom{1}0.6 && 0.08 && \rm{I} \\
 & \rm{SiO(5-4)} && \phantom{1}1.3 && 0.21 && \rm{I} \\
 \\
\object{DY Gem} & \rm{SiO(5-4)} && \phantom{1}1.1 && 0.09 && \rm{I} \\
 \\
\object{R Gem} & \rm{SiO(2-1)} && \phantom{1}0.2 && 0.06 && \rm{I} \\
 & \rm{SiO(3-2)} && \phantom{1}0.1 && 0.03 && \rm{S/B98} \\
 & \rm{SiO(5-4)} && \phantom{1}1.7 && 0.30 && \rm{I} \\
 \\
\object{RX Lac} &  \rm{SiO(2-1)} && \phantom{1}0.3 && 0.05 && \rm{I} \\
&  \rm{SiO(5-4)} && \phantom{1}1.7 && 0.30 && \rm{I} \\
&  \rm{SiO(8-7)} && \phantom{1}0.6 && 0.07 && \rm{J} \\
 \\
\object{Y Lyn} & \rm{SiO(2-1)} && \phantom{1}0.4 && 0.07 && \rm{I} \\
 & \rm{SiO(5-4)} && \phantom{1}2.6 && 0.33 && \rm{I} \\
 & \rm{SiO(8-7)} && \phantom{1}1.8 && 0.14 && \rm{J} \\
\\
\object{S Lyr} & \rm{SiO(2-1)} && \phantom{1}0.5 && 0.04 && \rm{I} \\
 & \rm{SiO(5-4)} && \phantom{1}3.0 && 0.19 && \rm{I} \\
\noalign{\smallskip}
\hline
\end{array}
$$
\end{table}
\begin{table}
\caption{Observational results of circumstellar SiO radio line emission (continued).}
\label{SiOintensities2}
$$
\begin{array}{p{0.15\linewidth}ccccccc}
\hline
\noalign{\smallskip}
\multicolumn{1}{l}{{\mathrm{Source}}} &
\multicolumn{1}{c}{{\mathrm{Transition}}} &&
\multicolumn{1}{c}{{I_{\rm{mb}}}} &&
\multicolumn{1}{c}{{T_{\rm{mb}}}} &&
 \multicolumn{1}{c}{\rm{Telescope}} \\ 
& &&
\multicolumn{1}{c}{\mathrm{[K \ km \ s^{-1}]}} &&
\multicolumn{1}{c}{[\mathrm{K}]} &&
\multicolumn{1}{c}{\rm{or \ Reference}}
\\
\noalign{\smallskip}
\hline
\noalign{\smallskip}
 \object{RT Sco} & \rm{SiO(3-2)} && 0.5 && 0.04 && \rm{S/B98} \\
 & \rm{SiO(8-7)} && 1.5 && 0.19 && \rm{A} \\
 \\
\object{ST Sco}  & \rm{SiO(3-2)} && 0.1 && 0.02 && \rm{S/B98} \\
& \rm{SiO(8-7)} && 0.5 && 0.07 && \rm{A} \\
 \\
\object{RZ Sgr}  & \rm{SiO(3-2)} && 0.3 && 0.03 && \rm{S/B98} \\ 
& \rm{SiO(8-7)} && 0.5 && 0.10 && \rm{A} \\
 \\
\object{ST Sgr}  & \rm{SiO(8-7)} && 0.6 && 0.08 && \rm{A} \\
 \\
\object{T Sgr} & \rm{SiO(8-7)} && 0.5 && 0.09 && \rm{A} \\
\\
\object{EP Vul} & \rm{SiO(2-1)} && 0.4 && 0.07 && \rm{I} \\
 & \rm{SiO(5-4)} && 1.8 && 0.27 && \rm{I} \\
 & \rm{SiO(8-7)} && 0.7 && 0.12 && \rm{A} \\
 \\
\object{DK Vul} & \rm{SiO(2-1)} && 0.2 && 0.03 && \rm{I} \\
 & \rm{SiO(5-4)} &&1.0 && 0.14 && \rm{I} \\
 \\ 
\object{AFGL 2425} & \rm{SiO(2-1)} && 0.9 && 0.07 && \rm{I} \\
 & \rm{SiO(5-4)} && 2.0 && 0.20 && \rm{I} \\
 & \rm{SiO(8-7)} && 0.5 && 0.08 && \rm{A} \\
 \\
\object{CSS2 41} & \rm{SiO(2-1)} && 1.0 && 0.05 && \rm{I} \\
 & \rm{SiO(5-4)} && 2.4 && 0.16 && \rm{I} \\
 \\
\object{IRC--10401}	& \rm{SiO(2-1)} && 2.9 && 0.14 && \rm{I} \\
 & \rm{SiO(5-4)} && 2.2 && 0.20 && \rm{I} \\
 & \rm{SiO(8-7)} && 1.4 && 0.12 && \rm{A} \\
 \\
\noalign{\smallskip}
\hline
\end{array}
$$
A; APEX, I; IRAM, J; JCMT, O; OSO, S; SEST  \\ B98; \citet{biegetal98}
\end{table}
\begin{table}
\caption{Observational results of circumstellar SiO($J$\,=\,2\,--\,1, $v$\,=\,1) emission.}
\label{SiOintensitiesm1}
$$
\begin{array}{p{0.15\linewidth}ccccccc}
\hline
\noalign{\smallskip}
\multicolumn{1}{l}{{\mathrm{Source}}} &
\multicolumn{1}{c}{{T_{\rm{mb}}}} &&
\multicolumn{1}{c}{{\sigma}} &&
\multicolumn{1}{c}{{T_{\rm{mb}}\times D^{2}}}  &&
\multicolumn{1}{c}{{\mathrm{prev.}}} \\
&
\multicolumn{1}{c}{[\mathrm{K}]} &&
\multicolumn{1}{c}{[\mathrm{K}]} &&
\multicolumn{1}{c}{[\mathrm{K \ pc^{2}}]^{\rm{a}}} &&
\multicolumn{1}{c}{{\mathrm{det.}^{\rm{b}}}} \\
\noalign{\smallskip}
\hline
\noalign{\smallskip}
\object{R And} & 0.52 && 0.016 && 4.7 \times 10^{4} && \mathrm{Y} \\	
\object{W And} & 0.72 && 0.027 && 5.7 \times 10^{4} && \mathrm{Y} \\
\object{VX Aql} & \cdots && 0.047 && 8.8 \times 10^{4} && \cdots \\
\object{W Aql} & 0.17 && 0.022 && 9.0 \times 10^{3} && \mathrm{Y} \\
\object{AA Cam} & \cdots && 0.022 && 4.2 \times 10^{4} && \cdots \\
\object{S Cas} & 1.9\phantom{1} && 0.025 && 3.7 \times 10^{5} && \mathrm{Y} \\
\object{V365 Cas} & \cdots && 0.026 && 3.0 \times 10^{4} && \cdots \\
\object{WY Cas}  & 0.15 && 0.014 && 5.4 \times 10^{4} && \cdots \\
\object{V386 Cep} & 0.70 && 0.019 && 1.5 \times 10^{5} && \cdots \\
\object{AA Cyg} & \cdots && 0.023 && 1.6 \times 10^{4} && \mathrm{N} \\ 		
\object{AD Cyg} & \cdots && 0.053 && 1.5 \times 10^{5} && \cdots \\
\object{R Cyg} & \cdots && 0.008 && 4.5 \times 10^{3} && \mathrm{Y} \\		
\object{$\chi$ Cyg}	& 8.1\phantom{1} && 0.082 && 9.8 \times 10^{4} && \mathrm{Y} \\
\object{TV Dra}	& 0.08 && 0.012 && 1.2 \times 10^{4} && \mathrm{N} \\
\object{DY Gem} & \cdots && 0.030 && 4.2 \times 10^{4} && \cdots \\
\object{R Gem} & 0.19 && 0.015 && 9.3 \times 10^{4} && \mathrm{N} \\
\object{ST Her} & \cdots && 0.022 && 6.0 \times 10^{3} && \cdots \\ 		
\object{RX Lac} &  \cdots && 0.032 && 9.3 \times 10^{3} && \cdots \\
\object{R Lyn} & \cdots && 0.027 && 6.0 \times 10^{4} && \mathrm{Y} \\
\object{Y Lyn} & \cdots && 0.012 && 2.4 \times 10^{3} && \mathrm{Y} \\		
\object{S Lyr} & \cdots && 0.037 && 1.6 \times 10^{5} && \cdots \\
\object{FU Mon} & \cdots && 0.038 && 6.9 \times 10^{4} && \mathrm{N} \\
\object{EP Vul} & \cdots && 0.050 && 3.9 \times 10^{4} && \mathrm{Y} \\
\object{DK Vul} & \cdots && 0.043 && 7.2 \times 10^{4} && \cdots \\
\noalign{\smallskip}
\hline
\end{array}
$$
$^{\rm{a}}$ For the non-detections we use 3$\sigma \times D^{2}$. \\
$^{\rm{b}}$ From \citet{joriknap98} and references therein.
\end{table}
%



\begin{figure*}[h]
\raggedright 
{\includegraphics[width=18cm]{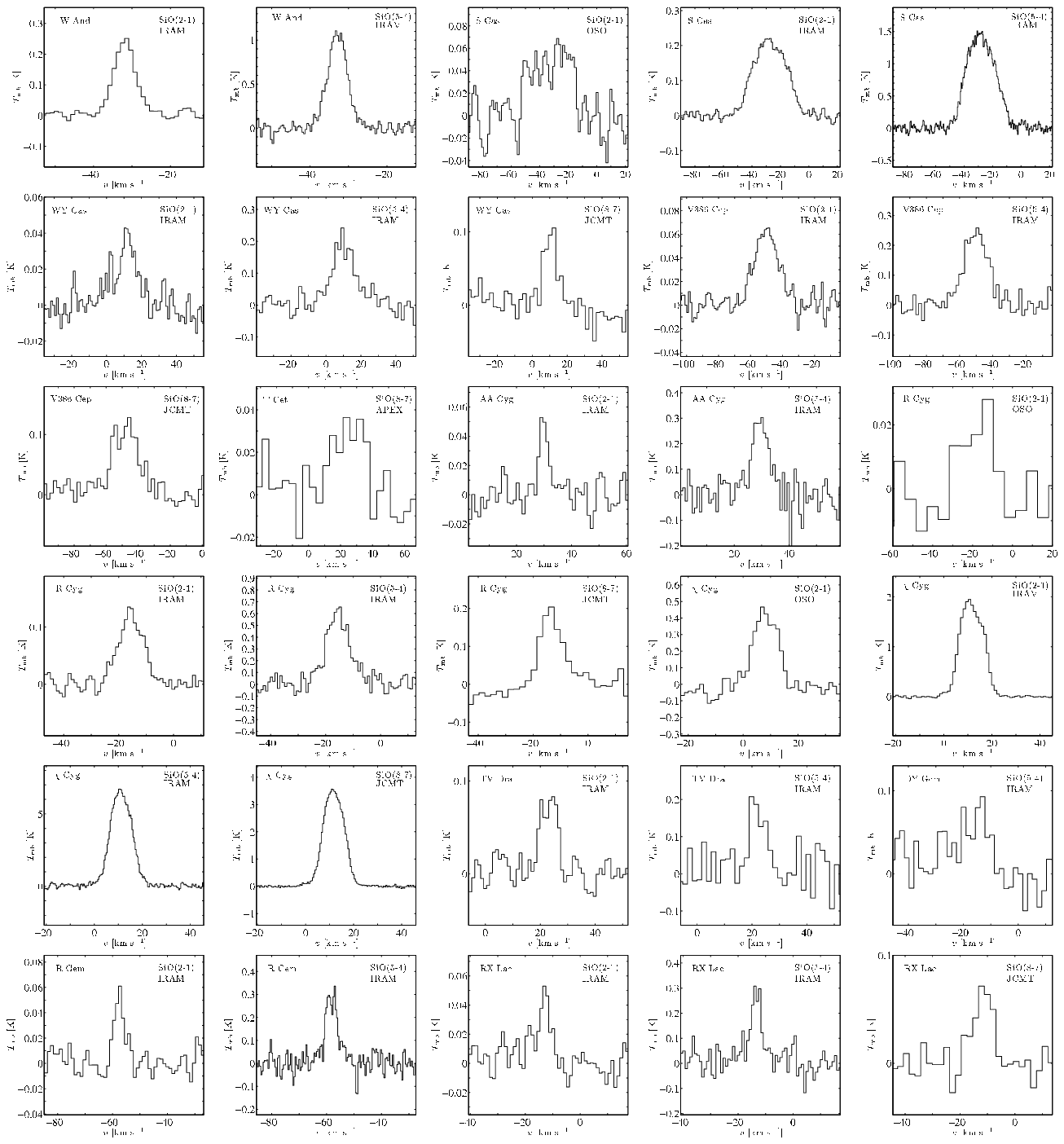}}
\caption{New observations of circumstellar SiO line emission ($v=0$). The source name is shown to the upper left, and the transition observed and the telescope used is shown to the upper right of each frame. LSR velocity scales are used.}
\label{rand_sio1}
\end{figure*}

\begin{figure*}[h]
\raggedright 
{\includegraphics[width=18cm]{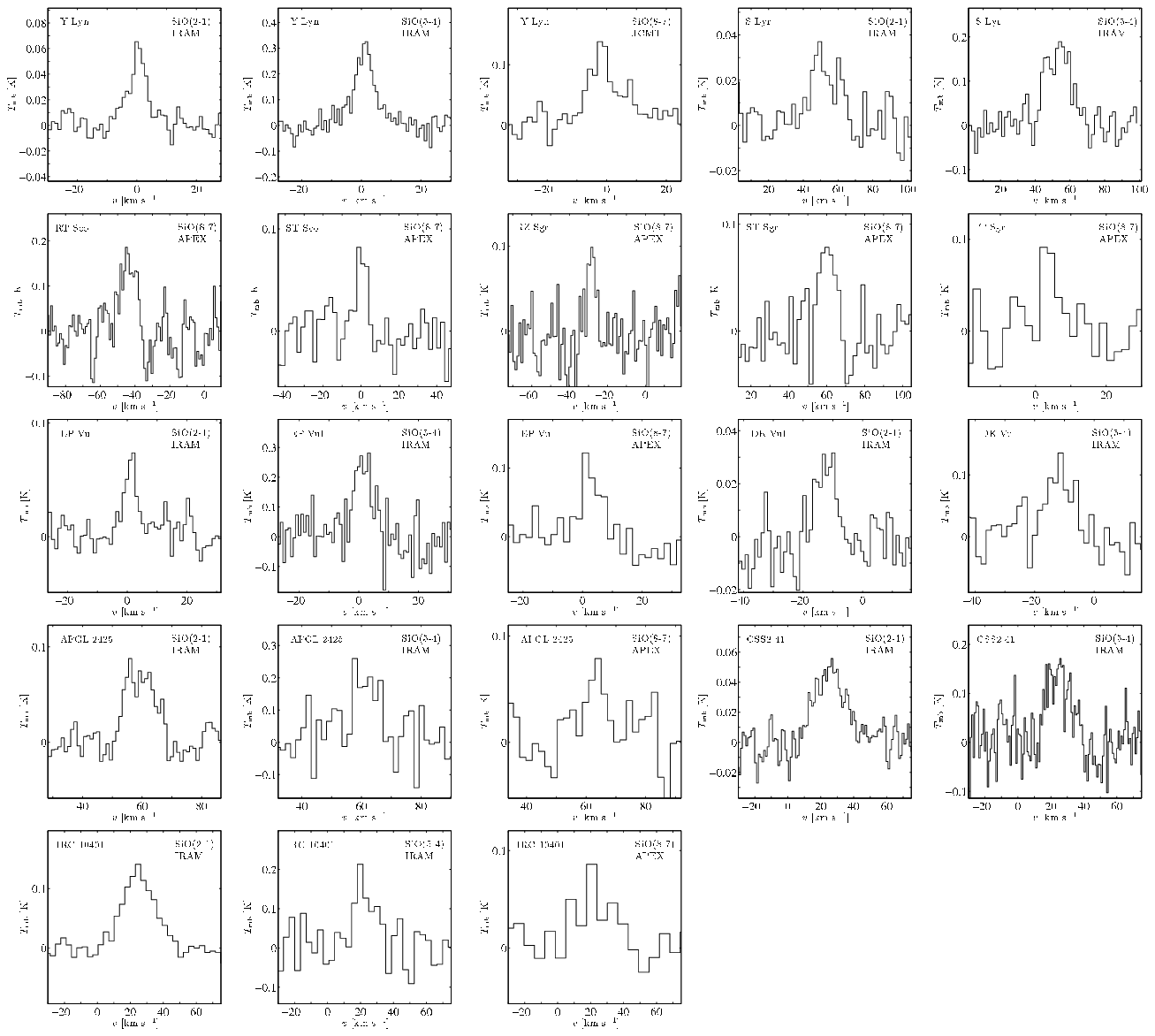}}
\caption{New observations of circumstellar SiO line emission ($v=0$). The source name is shown to the upper left, and the transition observed and the telescope used is shown to the upper right of each frame. LSR velocity scales are used.}
\label{rand_sio2}
\end{figure*}

\begin{figure*}[h]
\raggedright 
{\includegraphics[width=18cm]{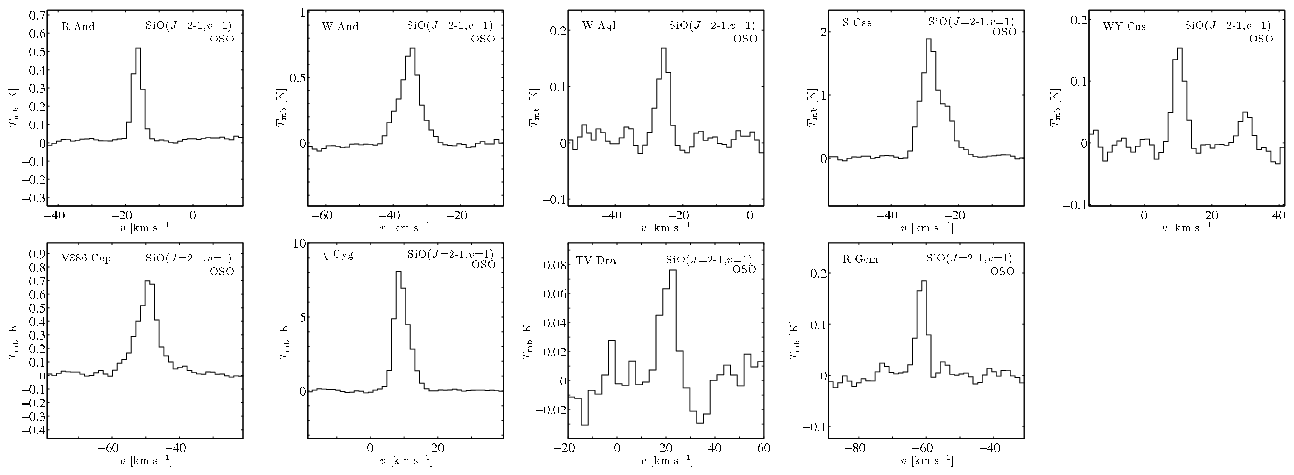}}
\caption{New observations of circumstellar SiO line emission ($v=1$). The source name is shown to the upper left, and the transition observed and the telescope used is shown to the upper right of each frame. LSR velocity scales are used.}
\label{rand_siom}
\end{figure*}

\end{document}